\documentclass[aps,floatfix,onecolumn,notitlepage,amsfonts,amsmath,amssymb,superscriptaddress,nofootinbib]{revtex4-1}

\usepackage[utf8]{inputenc}
\usepackage[T1]{fontenc}
\usepackage{accents}
\usepackage[normalem]{ulem}
\usepackage{graphicx}  

\usepackage{amsthm}
\newtheorem{theorem}{Theorem}

\newcommand{\ket}[1]{\left| #1 \right>}

\newcommand{\iprod}[2]{\langle #1 | #2 \rangle}

\newcommand{\oprod}[2]{| #1 \rangle\!\langle #2 |}

\newcommand{\vect}[1]{\mathbf{#1}}
\newcommand{\xvec}{\vect{x}}

\PassOptionsToPackage{hyphens}{url}
\usepackage[colorlinks=true,linkcolor=blue,urlcolor=blue,citecolor=blue,bookmarks=false]{hyperref}

\begin{document}

\title{Experimentally Probing the Algorithmic Randomness and Incomputability of Quantum Randomness}

\author{Alastair A.\ Abbott}
\affiliation{Univ.\ Grenoble Alpes, CNRS, Grenoble INP, Institut N\'eel, 38000 Grenoble, France}

\author{Cristian S.\ Calude}
\affiliation{Department of Computer Science, University of Auckland, Private Bag 92019, Auckland, New Zealand}

\author{Michael J.\ Dinneen}
\affiliation{Department of Computer Science, University of Auckland, Private Bag 92019, Auckland, New Zealand}

\author{Nan Huang}
\affiliation{Department of Computer Science, University of Auckland, Private Bag 92019, Auckland, New Zealand}

\date{\today}

\begin{abstract}
The advantages of quantum random number generators (QRNGs) over pseudo-random number generators (PRNGs) are normally attributed to the nature of quantum measurements.  
This is often seen as implying the superiority of the sequences of bits themselves generated by QRNGs, despite the absence of empirical tests supporting this.
Nonetheless, one may expect sequences of bits generated by QRNGs to have properties that pseudo-random sequences do not; 
indeed, pseudo-random sequences are necessarily computable, a highly nontypical property of sequences.

In this paper, we discuss the differences between QRNGs and PRNGs and the challenges involved in certifying the quality of QRNGs theoretically and testing their output experimentally.  
While QRNGs are often tested with standard suites of statistical tests, such tests are designed for PRNGs and only verify statistical properties of a QRNG, but are insensitive to many supposed advantages of QRNGs.  
We discuss the ability to test the incomputability and algorithmic complexity of QRNGs.  
While such properties cannot be directly verified with certainty, we show how one can construct indirect tests that may provide evidence for the incomputability of QRNGs.  
We use these tests to compare various PRNGs to a QRNG, based on superconducting transmon qutrits and certified by the Kochen-Specker Theorem, to see whether such evidence can be found in practice. 

While our tests fail to observe a strong advantage of the quantum random sequences due to algorithmic properties, the results are nonetheless informative: some of the test results are ambiguous and require further study, while others highlight difficulties that can guide the development of future tests of algorithmic randomness and incomputability.
\end{abstract}

\maketitle

\section{Introduction}

Randomness is an important resource in a diverse range of domains: it has uses in science, statistics, cryptography, gambling, and even in art and politics.
In many of these domains, it is crucial that the randomness be of high quality.
This is most directly the case in cryptography, where good randomness is vital to the security of data and communication, but is equally, albeit more subtly, true in other areas such as politics, where decisions of consequence may be made based on scientific and statistical studies relying crucially on randomness.

For a long time, people have predominantly relied on pseudo-random number generators (PRNGs)---that is, computer algorithms designed to simulate randomness---to serve such needs.  
Problems with various PRNGs, often only uncovered when it is already too late, are all too common and can have serious consequences.%
\footnote{An example is  the discovery in 2012 of a weakness in
the  encryption system used worldwide for online shopping, banking and email;  the flaw was traced to the numbers a PRNG had produced~\cite{factor_wrong2012}.  
As of 2018, Java still relies on a linear congruential generator, a low quality PRNG.} 
This has driven a recent surge of interest in RNGs exploiting physical phenomena, and more particularly in quantum RNGs (QRNGs) that utilise the inherent randomness in quantum mechanics~\cite{Svozil:1990aa,Stefanov:2000aa,Pironio:2010aa,Bera17}.  
QRNGs are generally considered to be, by their very nature, better than classical RNGs (such as PRNGs), but how (or can) one test this in practice?

In fact, as we will see, there are two distinct aspects of randomness we may consider when asking this question.
Firstly, one may look at the randomness of the process upon which a QRNG operates.
Testing this aspect of randomness purely from the output of a device (rather than \emph{a priori} considerations of its workings) is extremely difficult; classically it is not even possible and it is remarkable that, if one has access to an initial source of randomness, it is possible for certain QRNGs~\cite{Pironio:2010aa,Colbeck:2011gd}.
Nonetheless, such tests are extremely demanding, often require additional assumptions and are applicable only to QRNGs designed for such tests.

In practice, instead, one is restricted to studying the randomness of individual strings produced by QRNGs by, e.g., conducting batteries of tests on (finite) sequences they have produced~\cite{DIEHARD-paper,NISTtests}, and it is this problem that we concern ourselves with here.  
Although this is a different aspect of randomness than that of the process itself, QRNGs should also provide an advantage here: PRNGs are guaranteed to produce computable sequences in stark contrast to the incomputability of QRNGs~\cite{Abbott:2010fk,Abbott:2012fk,Calude:2008aa,Abbott:2015fk}.
Standard tests, however, have focused on intuitive aspects of randomness, such as the frequencies of certain (strings of) bits, but human intuition about randomness is notoriously poor~\cite{Bar-Hillel:1991aa,consciouns_rand20018} and many other symptoms of randomness remain untested.  
Indeed, the randomness of strings and sequences is an incomputable property and thus cannot be verified completely; moreover, it is characterised by an infinity of properties~\cite{Calude:2002fk}.  
One may wonder, however, whether there are tests more appropriate for analysing QRNGs and perceiving the advantage they can provide.

With this goal, we formulate and study several tests of randomness based on algorithmic information theory.  
In particular, we consider tests based on Borel normality~\cite{Borel09,Calude:1994fk} as well as novel tests based on the Solovay-Strassen probabilistic primality test~\cite{Solovay77,Calude:2010aa}---an algorithm which can be made deterministic when given access to algorithmic randomness~\cite{Chaitin78}.  
These latter tests allow one to probe indirectly the algorithmic randomness---and consequently also the incomputability---of sequences produced by a RNG, and thus have the potential to identify differences between QRNGs and PRNGs that are not captured by more traditional statistical tests.  
We test several classical RNGs as well as a semiconductor-based QRNG~\cite{PhysRevLett.119.240501} using these tests to see assess their utility for real RNGs.
While the first few tests we consider fail to find any significant difference between the quantum random sequences and those produced by the PRNGs, they bring to light certain issues useful for guiding future tests of algorithmic randomness and incomputability.
Our final test finds some significant differences between the QRNG and PRNGs, but it is unclear whether these are really due to algorithmic properties of the strings; limitations of this test mean that further study is needed.

\section{Randomness}

In order to guide the development of tests for QRNGs, it is important to understand what randomness is and thus what one should test.  
Historically, the quest to develop a formal understanding of randomness focused on the problem of determining whether a given (finite) string or (infinite) sequence of bits is random.  
One of the first attempts to formalise such a notion of randomness is due to Borel, who defined the concept of \emph{Borel normality} for infinite sequences~\cite{Borel09}.  
Borel normality formalises the notion that bits should be evenly and equally distributed within a sequence.  
Although this captures one of the most intuitive features of randomness, it does not alone
capture fully the desired concept.  
For example,  the \emph{Champernowne sequence} $0\,1\,00\,01\,10\,11\,000\,001\,011\,100\dots$ \cite{Champernowne:1933kx} contains every string of length $k$ with the same limiting frequency of $2^{-k}$, and yet the sequence has a simple description:
concatenate the binary representation of all the strings of length $k$ in
lexicographical order for $k=1,2,\dots$.  Given this description, it is clear
that the Champernowne sequence is not random, but highly ordered.

The study of algorithmic information theory, developed in the 1960s by
Solomonoff, Kolmogorov and Chaitin, provides more robust and acceptable
definitions of a random sequence. 
In this framework, random strings and sequences are those
that are incompressible~\cite{Chaitin:1977fk}. 
The incompressibility of strings depends on the choice of universal Turing machine; this shortcoming disappears when  the definition is extended to infinite sequences~\cite{Calude:2002fk,DH}.
Notions of randomness---both for finite strings and infinite sequences---defined  in terms of incompressibility are generically called \emph{algorithmic randomness}.

Let us briefly give some technical details useful later in the paper; we refer
the reader to~\cite{Calude:2002fk} for further details.  Consider Turing
machines operating on binary strings.  A  Turing machine $U$ is universal if
for every Turing machine $M$ there exists a prefix $p$  (depending only on $U$
and $M$) such that $U(px) = M(x)$, for every program $x$.  The
\emph{Kolmogorov} (or \emph{algorithmic}) \emph{complexity} of a Turing machine
$M$ is defined by $K_M(x) = \inf \{|s| : M(s)=x\}$, where by $|s| $ we
denote the length of the string $s$.  We can see  that $U$ is universal if and
only if for every Turing machine $M$ there exists a constant $c$ such that
$K_U(x) \le K_M(x)+c$, for every string $x$.  For this notion of complexity,
the running time and the amount of storage required for computation are
irrelevant.  One can prove that for every $M$ the maximum value of $K_M(x)$
over all strings $x$ of a fixed length $|x|=n$ is $n+\mbox{O}(1)$.
Furthermore, the overwhelming majority of strings $x$ of length $n$ have
$K_M(x)$ very close to $n$.  This means that almost all strings of length $n$
are incompressible by $M$: more formally, very few such strings have $K_M(x)<n$
(i.e., are compressible).  If $U$ is a universal Turing machine, then the
condition $K_M(x)< |x|$ means that $K_U(x)<  |x|-c$, that is, $x$ is
\emph{$c$-incompressible} (or \emph{$c$-Kolmogorov random}).  These
incompressible strings are highly random, patternless and typical.  It is easy
to prove that less than $2^{n-c}$ strings of length $n$ are not
$c$-incompressible. An infinite sequence 
$\xvec$  is called  \emph{Martin-L\"of random} if there exists a constant $C$ such that infinitely many prefixes of  $\xvec$  are $C$-Kolmogorov random. This definition  is equivalent to the condition that $\xvec$  
passes all  Martin-L\"of tests of randomness~\cite{Martin-Lof:1966kx}; see Section~\ref{mltest} for more details.

While algorithmic information theory provides a sound notion of randomness
for strings and sequences, two important points must be mentioned.  Firstly, it
is not effectively decidable whether a string or sequence is random, so the
notion does not provide a practical way to test the randomness of a finite or infinite sequence
of bits.  Secondly, it is possible to define ever stronger notions of
randomness: from an algorithmic perspective, no notion of ``true'', ``perfect" or
``absolute'' randomness exists, only degrees of
randomness~\cite{Calude:2002fk,Calude17,Graham:1990oe}.  This should temper any
desire to verify the randomness of a RNG by tests on its output.  Instead, we
can only hope to compare the quality of strings produced.

As interest in \emph{generating} random numbers soared, the concept of
randomness received increased philosophical attention and it became clearer
that the algorithmic notion of randomness fails to capture aspects of
randomness important for RNGs~\cite{Abbott15}.  Indeed, as von Neumann noted,
``there is no such thing as a random number---there are only methods to
produce random numbers''~\cite{Neumann:2012uq}.  The insight of von Neumann is
not that the algorithmic notion of randomness is problematic---indeed, it is
highly satisfactory as a notion of random \emph{objects}---but that there is a
dual concept of randomness, that of random
\emph{processes}~\cite{Eagle:2014jx,Abbott15,Solis15b}.  Such a concept has historically
received little attention, but the most convincing attempts to make it rigorous
are perhaps those which define it as a form of maximal unpredictability: the
outcome of such a process should be unpredictable for any physical
observer~\cite{Eagle:2005ys,Abbott:2015vg}.

The randomness of a process is often quantified in terms of entropy, but it is important to note that, entropy being a function of the probability distribution associated to a process, such a quantification requires i) knowing that the process is indeed unpredictable, and ii) knowing the probability distribution modelling its behaviour.
Although one can empirically estimate  the distribution from the output of a RNG, the entropy calculated from such data can only be interpreted as a measure of randomness if (i) is satisfied, and this cannot be directly verified from the empirical data alone.

There are thus two legitimate notions of randomness to be reconciled: that of
\emph{process randomness} (which is applicable to RNGs---viewed as processes---themselves),  and that of \emph{product randomness} (which is applicable to the strings---i.e.\ objects---obtained from RNGs).
The distinction between these notions is important for understanding tests of randomness.

\section{Random number generators (RNGs)}

An ideal random number generator is normally taken to be a random process
producing the same probability distribution as the ideal (but unphysical)
unbiased coin.  It thus produces bits sequentially, thereby generating a sequence
$\xvec=x_1x_2\dots$ with each bit $x_i$ being equiprobable, i.e.\
$p(x_i=0)=p(x_i=1)=1/2$, and with successive bits produced independently.
Hence, all strings $x$ of length $k$ have probability $p(x)=2^{-k}$ and, in the
infinite limit, one obtains the Lebesgue measure over all infinite
sequences~\cite{Calude:2002fk}.
This type of ideal source has maximal entropy.
It is important to recognise that this conception of an ideal RNG embodies the notion of random processes, not products, and concerns the distribution produced by said process and not its output.

If one tries to implement such a device in practice, two issues immediately become apparent.

Firstly, how is one to know that the process exploited is really random and
actually produces the expected ideal distribution?  This issue touches on the
interpretation of probability~\cite{Hajek:2014di} (although this is beyond the
scope of the present article).  For example, a physical process thought to be
represented by the uniform distribution might only exhibit epistemic
randomness, and a more precise, deterministic model of the process might be
possible which reveals its non-randomness.  
The most direct way to avoid such
possibilities is to harness an indeterministic process to ensure its
unpredictability~\cite{Abbott:2015vg}.

Secondly, how does one test or verify the randomness of a RNG given that one
only has access to (finite) strings produced by it?  Although the concepts of
process and product randomness are indeed distinct, they are nonetheless
related: long enough strings produced by an ideal RNG will, with high
probability, be incompressible, while in the infinite limit the sequences produced will
be Martin L\"of random (and thus also incomputable) with probability~1 \emph{but not with certainty}: an ideal
coin can in principle produce non-random or even computable sequences.
However, as mentioned earlier, the randomness of sequences is already an
incomputable property.  Thus, one can do no better than verifying finitely many
properties of randomness to gain confidence in a RNG.

\subsection{Pseudo RNGs (PRNGs)}

The predominant approach to generating randomness is to use algorithms to
produce ``pseudo-randomness'', and such PRNGs are ubiquitous as a result of
their practicality and speed.  However, the very fact that such devices use
computational methods to produce their outcomes distinguishes them from ideal
RNGs.  PRNGs typically use a short string from an external source---generally
assumed to be random---as an initial ``seed'' for an
algorithm~\cite{Gentle03}.  Thus, PRNGs can only produce computable sequences,
whereas such sequences should be produced only with probability~0 by an ideal
RNG.  Instead, effort is made to make PRNGs difficult to distinguish from an
ideal RNG given limited (typically polynomial time) computational
resources~\cite{Goldreich01}, so that the PRNG appears to be a high entropy source.
This provides a degree of security against
cryptographic attacks, even if the resulting distribution (induced by the
distribution over the initial seeds) is far from uniform in reality.

PRNGs generally produce sequences that satisfy many intuitive aspects of
randomness---such as the equidistribution of the bits produced---and pass
most standard statistical tests of randomness despite their computability.
Nonetheless, deficiencies resulting from the non-randomness of PRNGs are
regularly exploited (see, e.g.,~\cite{Bernstein13}) and much of the interest in
quantum randomness has been driven by the potential to avoid the shortcomings
of PRNGs.

\section{Quantum randomness}

For some time now, quantum mechanics has garnered interest as a potential
source of randomness for RNGs.  Such interest stems from the fact that certain
quantum phenomena, such as the radioactive decay of an atom or the detection of
a photon having passed through a beamsplitter, are generally taken to be
``intrinsically random'' under the standard interpretation of quantum
mechanics~\cite{Acin:2013qa}.  We will first discuss these claims in a little
more detail---since it is important to base the randomness of QRNGs on more
formal grounds rather than simply assuming such randomness---before discussing
one approach to the generation of quantum randomness in more detail.

Claims about quantum randomness originate with the fact that, as a formal
theory, quantum mechanics differs fundamentally from classical physics in that
not all observable properties are simultaneously defined with arbitrary
precision.  Instead, quantum mechanics, via the Born rule, only specifies the
probabilities with which individual measurement outcomes occur for the
measurement of a physical quantity---i.e., a quantum \emph{observable}.
Formally, if a system is in a quantum state $\ket{\psi}$ and one measures an
observable $A$ with spectral decomposition $A=\sum_i a_i P_i$, where we adopt
the notation $P_i=\oprod{i}{i}$ for rank-1 projection observables, then one
obtains outcome $a_i$ with probability
\begin{equation}
	P(a_i|\psi)=|\iprod{i}{\psi}|^2.\label{eqn:BornRule}
\end{equation}

Thus, whereas randomness in classical physics is due to incomplete knowledge of the
precise initial conditions of a system (e.g., as in chaotic systems)~\cite{Longo:2008ud}, 
in quantum mechanics it is intrinsic to the standard interpretation of the formal theory.

Nonetheless, the Born rule is a purely formal statement, and interpreting the
probability distribution specified by the Born rule remains the subject of
ongoing debate.  The orthodox interpretation, however, is that the distribution
should be understood ontically as representing an indeterministic
phenomenon~\cite{Acin:2013qa}.  Crucially, this interpretation is more than a
mere assumption: several well-known no-go theorems rule out classical
statistical interpretations of quantum randomness.

Bell's Theorem~\cite{Bell:1964fk} is the most well-known of these results, and
shows that a classical, local hidden variable theory cannot reproduce the
statistics of quantum correlations that are observed~\cite{Aspect:1982dp}
between entangled particles.  The Kochen-Specker Theorem~\cite{Kochen:1967fk},
although perhaps lesser known, pinpoints this breakdown in determinism in a more precise way: it shows that, for any quantum system with more than
2 dimensions, it is logically impossible to predetermine all measurement
outcomes prior to measurement in a noncontextual fashion (i.e., in a way which
is independent of other compatible---and thus non-disturbing---measurements
one can perform).

More recently, this theorem has been refined to show that the only observables
that can be predetermined in a noncontextual way are those for which the Born
rule assigns the probability 1 to a particular
outcome~\cite{Abbott:2013ly,Abbott:2015bx}.  More precisely, we say that an
observable $A$ is \emph{value definite} for a system prepared in a state $\ket{\psi}$ if it has a
predetermined measurement outcome $v_{\psi}(A)$.  The stronger result shows
that for systems of more than 2 dimensions, if we assume that any such value
definite observables should be noncontextual, then $A$ is value definite if and
only if $\ket{\psi}$ is an eigenstate of $A$; all other observables must be
\emph{value indefinite}.

This result makes the extent of quantum value indefiniteness---and thus
indeterminism---clear and pinpoints which measurements are protected by such
formal results. This not only allows some QRNGs to be based more rigorously  on
physical principles but also to  clarify the link between quantum randomness  and indeterminism.  Crucially, this
result also allows one to show that the measurement of such value indefinite
observables satisfies a strong form of unpredictability~\cite{Abbott:2015fk},
proving that one really cannot provide better predictions than the Born rule
specifies, and thus giving a stronger theoretical grounding to claims about the
form of quantum randomness proposed for QRNGs.

\section{Quantum RNGs (QRNGs)}
\label{sec:QRNG}

These properties of quantum measurements make them an ideal candidate for random
number generation: if one measures an observable for which the Born rule
predicts a uniform distribution, then the QRNG embodies a perfect coin.
Moreover, the results discussed above show that---subject to very reasonable
physical assumptions about how classical objects should behave---this
distribution cannot be given an epistemic interpretation and the corresponding measurement outcomes are thus truly of
indeterministic origin.  The attractiveness of QRNGs is further enhanced by the
possibility of obtaining high bitrates and the simplicity of their physical
models.  This is in contrast to RNGs based on classical physics, such as
chaotic systems.

Early QRNGs relied on features such as radioactive decay~\cite{Schmidt:1970aa},
but simpler systems based, for example, on measuring the
polarisation~\cite{Jennewein:2000ly,Stefanov:2000aa,Shen:2010vn} or detection
times~\cite{Stipcevic:2007fk} of photons, have become the norm due to the
practical advantages they provide.  Such approaches have led to the development
of commercial QRNGs, such as ID Quantique's Quantis~\cite{quantis-idQuantique}.

Many successful QRNGs exploit two-dimensional systems to generate randomness
(e.g.\ Quantis uses the polarisation of photons).  This greatly simplifies the
design and production of such devices but neither Bell's Theorem (which
requires entanglement) nor the Kochen-Specker Theorem (which requires  at least
3-dimensional systems) are applicable, and these QRNGs thus lack the rigorous
theoretical certification that quantum mechanics can provide, even if it may be
reasonable to think that the measurements they exploit should still be
indeterministic.

The most direct approach to overcoming this shortcoming is to use higher dimensional systems for which the value indefiniteness of measurement outcomes is, via the Kochen-Specker theorem, provable~\cite{Abbott:2010fk,Abbott:2012fk} to certify a QRNG.
Such certification is necessarily ``device dependent''---that is, it relies on knowledge of the functioning of the QRNG---but nonetheless allows the randomness of the device to be more formally grounded.
A simple example of a QRNG certified in this way was proposed in~\cite{Abbott:2010fk,Abbott:2012fk} for spin-1 particles, but  its principle is applicable to any 3-dimensional system (i.e., an implementation of a qutrit). 
The approach proposed was to prepare a qutrit in the state $\ket{0}$ before measuring the observable $A=a_0\oprod{0'}{0'}+a_1\oprod{1'}{1'}+a_2\oprod{2'}{2'}$ for which the orthonormal basis $\{\ket{0'},\ket{1'},\ket{2'}\}$ is chosen such that $\iprod{0}{0'}=0$ and $\iprod{0}{1'}=\iprod{0}{2'}=\frac{1}{\sqrt{2}}$ (see Figure~\ref{fig3:realisation} below).  
Since the state $\ket{0}$ is thus an eigenstate of the projection observable $P_{0'}=\oprod{0'}{0'}$, this observable is value definite with value $v(P_{0'})=0$---that is, the measurement outcome $a_0$ never occurs.%
\footnote{This is, of course, only true in the ideal case. In the non ideal scenario, any such outcomes can simply be discarded.}
However, by the results of~\cite{Abbott:2012fk,Abbott:2015bx}, both $P_{1'}=\oprod{1'}{1'}$ and $P_{2'}=\oprod{2'}{2'}$ are value indefinite and, moreover, both outcomes $a_1$ and $a_2$ occur with probability $1/2$ according to the Born rule~\eqref{eqn:BornRule}.  
Thus, the QRNG operates as an ideal coin certified by value indefiniteness.

A QRNG based on this proposal has recently been implemented experimentally~\cite{PhysRevLett.119.240501}, not with spin-1 particles but by exploiting a superconducting transmon coupled to a microwave cavity as a qutrit.  
Figure~\ref{fig3:realisation} shows a schematic of the QRNG proposed in~\cite{Abbott:2010fk,Abbott:2012fk} based on the implementation used by Kulikov et al.~\cite{PhysRevLett.119.240501}.  
This implementation was used to generate a large number of bits, and in the subsequent sections we will analyse sample sequences produced by this QRNG implementation.  
In particular, we will look to detect differences between such sequences and pseudo-random sequences arising from algorithmic properties of the sequences.

\begin{figure}[t]
 \begin{center}
 \includegraphics[width=\textwidth]{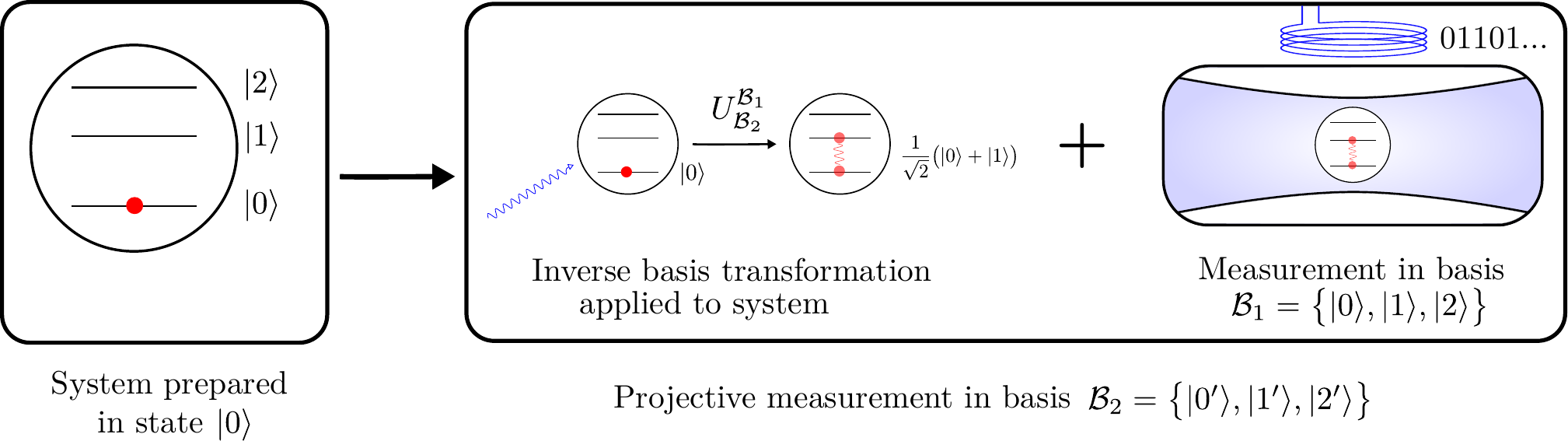}
 \caption{Schematic showing the QRNG based on the Kochen-Specker Theorem as implemented in~\cite{PhysRevLett.119.240501}. A transmon qutrit system is initially prepared in the state $\ket{0}$ (with respect to the computational basis $\mathcal{B}_1=\{\ket{0},\ket{1},\ket{2}\}$ by thermal cooling. The system is then measured in the basis $\mathcal{B}_2=\{\ket{0'},\ket{1'},\ket{2'}\}$ with $\iprod{0}{0'}=0$ and $\iprod{0}{1'}=\iprod{0}{2'}=\frac{1}{\sqrt{2}}$. In practice, this measurement is performed by first performing the inverse basis transformation on the system and measuring in the basis $\mathcal{B}_1$. Since $|\iprod{0}{0'}|^2=0$, this outcome never occurs in an ideal implementation, so the outcomes $a_1$ and $a_2$ corresponding to $\oprod{1'}{1'}$ and $\oprod{2'}{2'}$ are mapped to a binary sequence.}
\label{fig3:realisation}
\end{center}
\end{figure}

This approach to certifying a QRNG via value indefiniteness implies some additional interesting algorithmic properties of the output sequences of the device if one is willing to accept slightly
stronger physical assumptions (in particular, about whether being able to
compute properties in advance implies well-defined physical properties).
Specifically, it was shown in~\cite{Abbott:2012fk} that such a device, if used
repeatedly \emph{ad infinitum} to generate an infinite sequence $\xvec$ of
bits, will produce a sequence that is strongly incomputable (technically,
``bi-immune''~\cite{DH}) not just with probability 1, but \emph{with certainty}.  Although
such a result will not alone lead to observable advantages  for finite strings
-- recall that, from the Born rule, an ideal QRNG will produce an incomputable sequence with
probability 1---this nonetheless highlights the differences between pseudo and
quantum randomness in relation to computability.

More recently there has also been growing interest in a different type of QRNG which can provide a stronger form of certification but requires initial random seeds as input. (Such devices are thus technically randomness expansion devices, rather than RNGs.)
Typically, such devices rely on violating a Bell inequality, which allows one to certify that the QRNG indeed uses a value indefinite system without assuming \emph{a priori} anything about the workings of the device~\cite{Pironio:2010aa,Colbeck:2011gd,Ma16}. 
This type of certification is thus termed ``device independent'', and allows one to place lower bounds on the entropy of the source; it is particularly important in cryptographic settings, where one perhaps does not wish to trust the workings of a given RNG. 
Such schemes are very costly, however: not only is an initial random seed required, but one also must separate the QRNG into two space-like separated (or at least isolated) components and the stringent requirements of loophole-free Bell tests
reduce the obtainable bitrate by several orders of magnitude compared to
``standard'' QRNGs~\cite{Pironio:2010aa}.

Other related randomness expansion schemes have also been proposed which are less experimentally demanding but require additional physical assumptions~\cite{Himbeeck17}.
In particular, we note that ``noncontextuality inequalities''~\cite{Klyachko08,Cabello08} obtainable from proofs of the Kochen-Specker Theorem can be used to provide such a certification~\cite{Um:2013tt,Deng13,Miller17}.
In doing so, rather than trusting outright that the QRNG uses a system in which the Kochen-Specker Theorem applies, one actively verifies this under weaker physical assumptions about the workings of the device. 
Nonetheless, such schemes are still significantly more demanding than that described in Fig.~\ref{fig3:realisation}.
Here we thus focus on the similar device-dependent type scheme described in detail above. Indeed, our focus is on testing the algorithmic properties of individual strings produced by a QRNG; such tests are complementary to those aimed at certifying the indeterministic nature of the process itself, and the simplicity of this scheme, along with its high bitrate, facilities such an analysis.

\section{Testing RNGs}\label{sec:testingRNGs}

While it is crucial to have a good theoretical understanding of any RNG, there
are several reasons why testing experimentally their outputs is nonetheless
crucial.  Firstly, one can never be sure that the implementation of a RNG
matches its theoretical description, a fact that is equally as true for hardware RNGs as for
software RNGs.  Indeed, in the extreme limit, one might not wish to trust any
theoretical claims about a given RNG, and thus confidence in the RNG can only
be gained from performing carefully selected tests.  Secondly, thorough testing
gives one the opportunity to detect any issues with assumptions made in the
theoretical analysis of a device or in its practical deployment (e.g., if the
distribution of seeds does not match that assumed theoretically the performance
of a RNG might be compromised).

It is nonetheless important to recognise that experimental testing can never
allow one to perfectly characterise a device.  Instead, with access to only
finite strings produced by it and the ability to perform a finite number of
tests, one can only ever gain increasing confidence in the operation of the
device.  One can never be certain, for instance, that the output obtained was not a
simply atypical behaviour obtained purely by chance.  This is doubly true
since, as we discussed earlier, randomness is characterised by an infinity of
properties, so one must carefully choose the tests one performs.

The issues arising when testing the outputs of RNGs can be illustrated pointedly with an
example.  Imagine a device which deterministically outputs the digits of the
binary expansion of $\pi=\pi_1\pi_2\pi_3\dots$ starting from the $10^{10}$th
bit.  If we are unaware of the behaviour of this device and believe it to be a
RNG, its output will appear extremely random to us; indeed, $\pi$ passes all
standard statistical tests of randomness~\cite{Marsaglia05} despite the fact
that it is not even known to be Borel normal~\cite{Wagon04,MR3004253}.  Nevertheless, the
sequence produced by this box would be computable and thus not random at all.
Similarly, any attempt to estimate the entropy of the source from the empirically observed distribution would lead one to believe the source to be a highly random process, despite the fact it is completely deterministic.

Standard statistical tests of randomness focus on properties of the
distribution of bits or bit strings within sequences, properties more closely
related to Borel normality than algorithmic complexity.  Many such
tests were developed with the aim of testing PRNGs, where reproducing such
statistical predictions is a primary issue, particularly since failing to do so
may leak information about the seed and thus break the security of the
PRNG~\cite{factor_wrong2012}. QRNGs\footnote{As we discussed in the previous section, we restrict our discussion henceforth to standard QRNGs, rather than device-independent randomness expansion schemes. These devices remain the standard approach to QRNGs in practice, and developing tests for their output remains a crucial problem despite the increased interest in device-independent QRNGs.} have generally been tested against similar
tests, such as the NIST~\cite{NISTtests} and DIEHARD~\cite{DIEHARD-paper}
batteries, and generally perform well.  For example, Quantis is officially
certified as passing these tests on 1000 samples of 1 million
bits~\cite{quantis-idQuantique}.  Such tests, however, far from confirm the
randomness of the device; indeed, analysis of longer sequences (of $2^{32}$
bits) revealed (albeit it very small) bias and correlation amongst the output
bits~\cite{Abbott:2014rt}.

Such statistical non-uniformity is, however, to be expected in RNGs exploiting
physical phenomena due to experimental imperfections and
instability~\cite{Abbott:2010fk}. Inasmuch as this form of non-uniformity is
small enough for the required application, this is not necessarily problematic
as long as a QRNG remains certified by value indefiniteness: unlike for PRNGs,
where non-equidistribution is often a symptom of deeper issues, the unpredictability
of QRNGs is a result of the indeterministic nature of the device, and is thus
assured even if the resulting distribution is biased~\cite{Abbott:2015fk}.
Moreover, bias can be reduced by careful 
post-processing~\cite{Neumann:2012uq,Peres:1992aa,Abbott:2010ij}, allowing quantum
indeterminism to still be exploited sufficiently.  Although testing such
properties is crucial in order to ensure any bias remains tolerably low, such
tests do not directly probe crucial advantages of quantum randomness, such as the
degree of algorithmic randomness or incomputability of their output.

Some authors have also looked at the compressibility of quantum random sequences using standard compression algorithms~\cite{Kovalsky18} as a proxy for direct tests of Kolmogorov complexity.
In practice, however, just like the aforementioned tests, this approach fundamentally relies on statistical properties of the sequence and suffers from similar problems as the above tests (such as being fooled by computable sequences).
Indeed, it is not possible to directly compute the Kolmogorov complexity since it is an incomputable quantity.
Nevertheless, one may still ask whether there are  useful tests
that indirectly probe this to try and differentiate PRNGS---which always
produce computable sequences---from QRNGs~\cite{Calude:2010aa}.  In the following sections we
investigate more closely this question.

\section{Experimentally testing for evidence of incomputability and algorithmic randomness}

In this section we describe several tests based on algorithmic properties which
we use to study random bits obtained from both PRNGs
and the QRNG detailed in Figure~\ref{fig3:realisation}.  We tested 80 sequences
of $2^{26}$ bits\footnote{The sequences were obtained from 10 longer sequences of $2^{29}$ bits, each obtained during separate experimental runs. We split them further into smaller sequences in order to provide a more detailed statistical analysis.} obtained from each of the following six sources: the initial
bits of the binary representation of $\pi$ (which can be seen as a form of
pseudo-randomness~\cite{MR3004253}), the PRNG used by Python v3.5.4 (a Mersenne Twister
algorithm)~\cite{Matsumoto98}, Random123 v1.09~\cite{Random123},  PCG v0.98~\cite{PCG},
xoroshiro128+~\cite{xoroshift}, and the QRNG described in Section~\ref{sec:QRNG}
(see~\cite{PhysRevLett.119.240501}).

\subsection{Tests of Borel normality}
\label{sec:BN}

As mentioned earlier, the notion of Borel normality was the first mathematical
definition of algorithmic randomness~\cite{Borel09}, and although, like many
standard tests of randomness, it focuses on the distribution of bits within a
sequence, it is nonetheless worth looking at in its own right.

An infinite sequence $\xvec\in\{0,1\}^{\infty}$ is (Borel) normal if every
binary string appears in the sequence with the right frequency (which is
$2^{-n}$ for a string of length $n$).  Every Martin-L\"of random infinite
sequence is Borel normal~\cite{Calude:2002fk}, but the converse implication is
not true: there exist computable normal sequences, such as Champernowne's
sequence mentioned earlier.  Normality is invariant under finite variations:
adding, removing, or changing a finite number of bits in any normal sequence
leaves it normal.

The notion of normality was subsequently transposed from infinite sequences to
(finite) strings~\cite{Calude:2002fk}.  In doing so, one has to replace limits
with inequalities, and one obtains the following definition.  For any fixed
integer $m > 1$, consider the alphabet $B_{m} = \{0,1\}^{m}$ consisting of all
binary strings of length $m$, and for every $1 \leq i \leq 2^{m}$ denote by
$N_{i}^{m}$ the number of occurrences of the lexicographical $i$th binary
string of length $m$ in the string $x$ (considered over the alphabet $B_{m}$).
By $|x|_{m}$ we denote the length of $x$ over $B_m$; $|x|_1=|x|$.  A string
$x\in B_m$ is \emph{Borel normal} (\emph{with accuracy} $\frac{1}{\log_2|x|  }$) if
for every integer $1 \leq m \leq \log_2  \log_2   |x|$ and  each $1 \leq j \leq
2^{m}$ we have:
\begin{equation}
	\label{eq:inormality}
	\left| \frac{N_{j}^{m}(x)}{|x|_{m}} -  2^{-m} \right| \leq\frac{1}{\log_2  |x|}\raisebox{0.5ex}{.}
\end{equation}
Almost all algorithmic random strings are Borel normal with accuracy $\frac{1}{\log_2 |x| }$ \cite{Calude:2002fk}; in particular, they have approximately the same number of 0s and 1s.
Furthermore, if all prefixes of a  sequence are  Borel normal, then the sequence itself is also  Borel normal.

\begin{figure}[t]
 \begin{center}
 \includegraphics[width=0.6\textwidth]{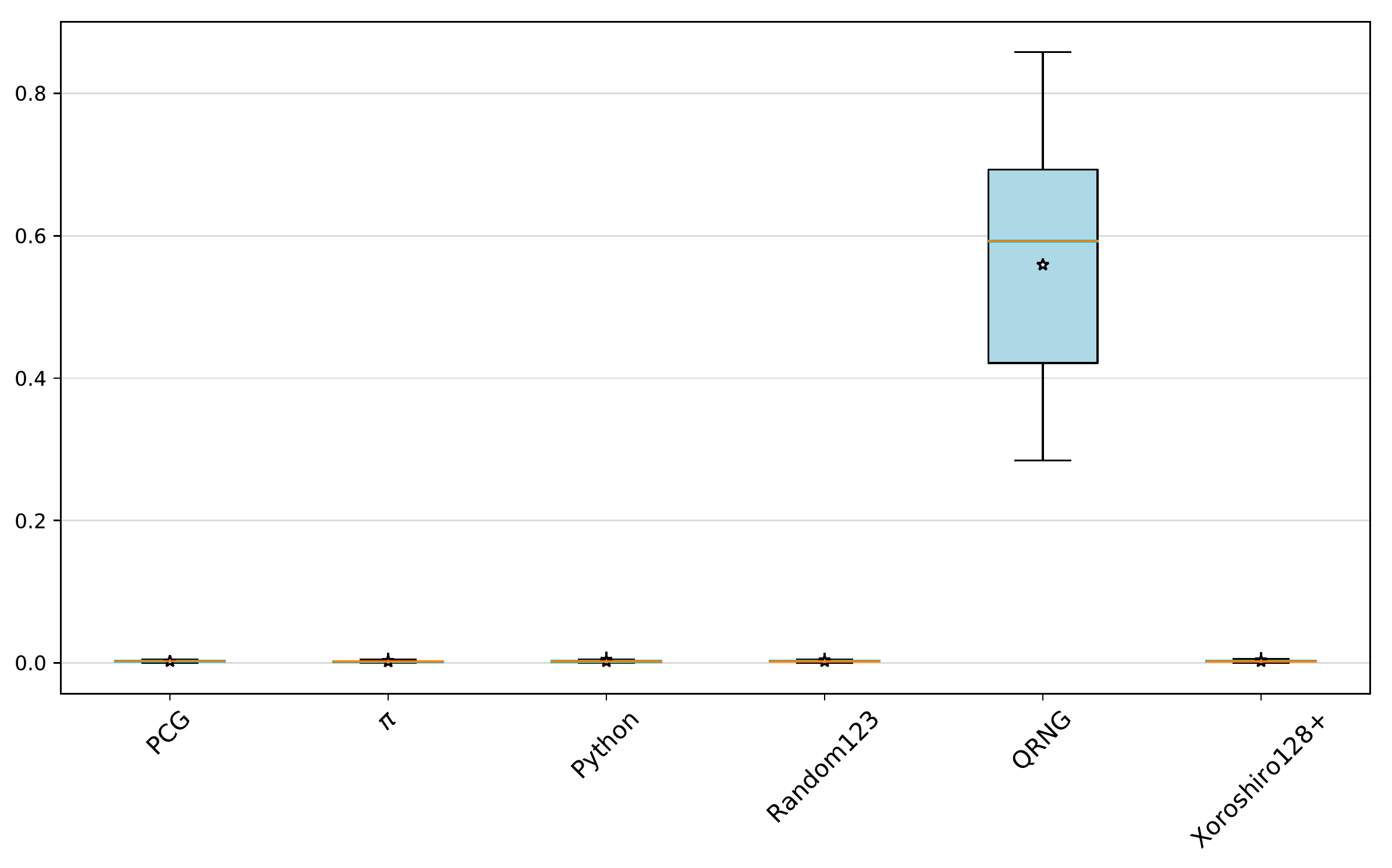}
 \caption{Borel normality test: Box-plot showing the distribution of the quantity $\operatorname{max}\left(\left| \frac{N_{j}^{m}(x)}{|x|_{m}} -  2^{-m} \right|\right) \log_2  |x|$ for the 80 strings of length $|x|=2^{26}$ bits produced by each the six RNGs tested.}
\label{fig:borelNormal}
\end{center}
\end{figure}

The fact that Borel normality for finite strings is only defined up to the accuracy function arises from the fact that the definition is well behaved (and converges to the definition for sequences in the limit) if the right-hand-side of Eq.~\eqref{eq:inormality} is replaced by any decreasing computable real function in $|x|$ converging to 0.
Fixing a specific accuracy function allows one to test explicitly the normality of finite sequences (and such tests have previously been performed on strings produced by QRNGs~\cite{Calude:2010aa,Solis15,Martinez18}), but such a choice of accuracy function is necessarily somewhat arbitrary.
However, the relative normality of strings can be tested by  comparing the values of a metric based on~\eqref{eq:inormality}; a reasonable choice of such a metric is the quantity $\operatorname{max}\left(\left| \frac{N_{j}^{m}(x)}{|x|_{m}} -  2^{-m} \right|\right) \log_2  |x|$ over the values $m=1,\dots, \lfloor\log_2\log_2|x|\rfloor$ and each $1\le j \le 2^m$.
We recorded this metric for the six sources of random bits under consideration, and the resulting distributions are shown in Figure~\ref{fig:borelNormal}.

The results show clearly that the bits produced by the QRNG are significantly
less normal than those produced by the other sources.  This is, however,
not surprising, since the experiment implementing the QRNG was known to exhibit bias
due to experimental imperfections~\cite{PhysRevLett.119.240501}.
Although it is possible to use normalisation procedures to unbias a source, simple techniques significantly reduce the length of the output strings (and thus the obtainable statistical power) and can alter various computability theoretic properties of a sequences~\cite{Abbott:2010ij}; conversely, the consequences of more complicated techniques on such properties of the output are poorly understood, so we opted against performing any such normalisation.
Nonetheless, as discussed at the end of Section~\ref{sec:testingRNGs}, a sufficiently small bias
may be less problematic in practical applications for QRNGs than for
traditional PRNGs.

While examining the normality of sequences produced by any RNG is  important,
this algorithmic property fails to test properties of algorithmic randomness or
incomputability in the way we aim to do.  The example of Champernowne's
sequence again testifies to this.  To probe such behaviour of QRNGs we
thus need to delve further into algorithmic properties of randomness.

\subsection{A Martin-L\"of test of incomputability}
\label{mltest}

Is it possible to give formally a test which rejects every
computable sequence as nonrandom?  
Martin-L\"of randomness is an important, if not the most important, form of algorithmic randomness
and is based on the notion of Martin-L\"of test of randomness. 
A test of randomness is  defined by a uniformly computably enumerable shrinking sequence of constructive open sets in Cantor space (the components of the test) whose intersection is a constructive null set (with respect to Lebesgue measure); see~\cite{Calude:2002fk} for more details. 
A sequence passes the test if it is not contained in this null set. A sequence is Martin-L\"of random if it passes all Martin-L\"of tests.
There exist countably many such tests: some test normality, others test  the law of large numbers, etc.
The answer to the question above is affirmative: such a Martin-L\"of test exists.

Testing incomputability rather that randomness directly is an important initial step: indeed, all algorithmically random sequences are incomputable and it is this property of randomness that PRNGs fail most starkly, since they necessarily produce computable sequences. Moreover, the robustness of incomputability to bias in a sequences makes such tests potentially more robust in practice.
To specify a Martin-L\"of test for computability, we must define the sequences contained in its $n$th
component for all integers $n>0$.  
To do so, one can take the $n$th component to be the union of all $\sigma \{0,1\}^{*}\{0,1\}^{\infty}$ for which there is an $e$ such that
$\sigma(0)=\varphi_e(0), \dots , \sigma(e+n+1) = \varphi_e(e+n+1)$ and $\sigma
\in \{0,1\}^{*}$.  This is an open computably enumerable class that contains
all computable sets, as each computable set has a computable characteristic function
$\varphi_e$.  Furthermore, the measure of the $n$th component is bounded from
above by $ \sum_e 2^{-n-e-2}$, which in turn is bounded from above by
$2^{-n-1}$, as the string $\sigma$ derived from $\varphi_e$ has length $e+n+2$ and is
a prefix of the set for which $\varphi_e$ computes the characteristic function.

It is not difficult to see that the above test for computability depends on the enumeration
$(\varphi_e)$, and there is no obvious ``natural'' choice.  Furthermore,
invariance under finite variations renders the test unsuitable for finite
experiments.  As a result, it is necessary to consider more indirect methods to
test the incomputability of sequences produced by RNGs.

\subsection{Chaitin-Schwartz-Solovay-Strassen tests}
\label{CSSStests}

In this section we propose and carry out several related tests based on a
rather different property of random sequences: their ability to de-randomise
the Solovay-Strassen probabilistic test of primality~\cite{Solovay77}.  In
contrast with most standard tests of randomness which check specific properties
of strings of bits, these tests are based on the behaviour of the strings with
respect to certain ``secondary'' tasks.  We first briefly describe the
Solovay-Strassen primality test and the advantage offered in this task by
random strings, before presenting the tests themselves.

The Solovay-Strassen test checks the primality of a positive integer $n$: take $k$ natural numbers uniformly distributed between $1$ and $n -
1$, inclusive, and, for each $i(=i_1,\dots,i_k)$, check whether a certain, easy to compute,
predicate $W(i, n)$ holds ($W$ is called the Solovay-Strassen predicate).  If $W(i,
n)$ is true then ``$i$ is a witness of $n$'s compositeness'',  hence $n$ is
composite.  If $W(i, n)$ holds for at least one $i$ then $n$ is composite;
otherwise, the test is inconclusive, but in this case the probability that $n$
is prime is greater than $1-2^{-k}$.  This is due to the fact that \emph{at
least half} the $i$'s between $1$ and $n - 1$ satisfy $W(i, n)$ if $n$ is 
composite, and \emph{none} of them satisfy $W(i, n)$ if $n$ is
prime~\cite{Solovay77}.

Chaitin and Schwartz~\cite{Chaitin78} proved that, if  $c$ is a large enough positive integer and  $s$ is a long enough
$c$-Kolmogorov random binary string, then $n$ is prime if and only if  $Z(s,n)$ is true,
where $Z$ is a predicate constructed directly from $O(\log n)$ conjunctions of
negations of $W$ predicates (see Section~\ref{sec:CSSStest3} below for more details).  
The crucial fact is that the set of $c$-Kolmogorov random
strings is highly incomputable: technically the set is immune, that is, it
contains no infinite computably enumerable subset~\cite{Calude:2002fk}.  As a
consequence, de-randomisation is thus non-constructive, and thus without practical value.

Drawing on this result, we propose several tests that operationalise it
in order to test the randomness of a sequence based on whether certain numbers obtained from RNGs succeed
in witnessing the compositeness of well chosen targets.  
We will make particular use of Carmichael numbers as these target composites.
A Carmichael number is a composite positive integer $n$ satisfying the congruence $b^{n-1} \equiv 1
\pmod n$ for all integers $b$ relatively prime to $n$.  Although Carmichael numbers are
composite, they are difficult to factorise and thus are ``very similar'' to
primes; they are sometimes called pseudo-primes. 
Many Carmichael numbers can pass
Fermat's primality test, but less of them  pass the Solovay-Strassen test. Increasingly
Carmichael numbers become ``rare''.\footnote{There are 1,401,644 Carmichael
numbers in the interval $[1, 10^{18}]$.}

In what follows we thus present four different tests based on the Chaitin-Schwartz
Theorem and the Solovay-Strassen test.
Since the proposed  tests rely directly on the
algorithmic randomness of a string, they can potentially give direct empirical
evidence of incomputability, in stark contrast to most tests of randomness.
For example, the Borel normality test discussed previously is unable to do so: the
normality of Champernowne's sequence mentioned earlier is evidence of this.

Since the Chaitin-Schwartz Theorem relies on the Kolmogorov randomness of the sequences it involves, these tests also go beyond the previous one in not only probing incomputability, but also algorithmic complexity more generally.
Indeed, an ideal QRNG should produce $c$-Kolmogorov random strings with very high probability, while PRNGs produces strings of very low Kolmogorov complexity (since, in the limit, they are computable).
Nonetheless, we focus on probing the incomputability of strings from QRNGs rather than their Kolmogorov complexity or randomness, a doubly motivated choice.
Firstly, the fact that incomputability is a weaker property than Kolmogorov randomness and less affected by bias means that any difference between pseudo and quantum randomness will potentially be easier to observe.
Secondly, as mentioned earlier, subject to an additional physical assumption, QRNGs can be shown to produce incomputable sequences with certainty, and not just probability one~\cite{Abbott:2012fk}.

As in~\cite{Calude:2010aa}, we conduct various statistical tests to determine whether any observed difference is statistically significant or not.
If a difference is found to be significant, we then look at whether this really provides evidence of incomputability or not.
As it is not \emph{a priori} clear what distribution the various test metrics we employ should follow,
we utilise the non-parametric and distribution free Kolmogorov-Smirnov test for
two samples~\cite{Conover} to determine whether two datasets differ
significantly.  This test returns a $p$-value\footnote{Exact $p$-values are
only available for the two-sided two-sample tests with no ties.} indicating the probability, given the observed test statistic, that the observed distributions were indeed drawn from the same distribution. 
We conclude that ``the difference between the two datasets is
statistically significant'' if the $p$-value is less than $0.005$. 
We choose this relatively strict $p$-value to lower the chance of false positives arising from the fact that we will perform several tests between several different data sources:
the probability of observing a spurious difference (simply by chance) on at least one of the many tests is much higher than the critical $p$-value of $0.005$ of obtaining such a spurious result on any single test. A higher critical $p$-value (such as the commonly used $0.05$) would mean such false positives would be highly probable.

When no significant difference is found by the Kolmogorov-Smirnov test, we additionally check whether the test metric distribution is consistent with a normal distribution by performing a Shapiro-Wilk test~\cite{Shapiro-Wilk};\footnote{More precisely, the Shapiro-Wilk test examines the null hypothesis that the samples $z_{1},\ldots
,z_{n}$ come from a normally distributed population. This test is appropriate
for small samples, since it is not an asymptotic test.} if it is,\footnote{Here we consider evidence for non-normality to be a $p$-value below $0.05$.} we then use the (parametric) Welch $t$-test~\cite{Welch}, which is a version of Student's test, to determine whether there is a significant difference between the means of the test statistics for the different RNGs under the assumption of normally distributed test metrics.

\subsubsection{First Chaitin-Schwartz-Solovay-Strassen test}

The first test we look at, which was previously used in~\cite{Calude:2010aa},
probes directly the efficacy of a set of random bits in simulations (in our
case for checking primality).

We performed this test on  all of the  246,683 Carmichael numbers $n$ with at most 16 digits
as computed in~\cite{Pinch07}, using strings of bits from each random source to
specify the numbers tested as potential witnesses of compositeness.  More
precisely, for a fixed $k$ (see below) and each Carmichael number $n$ we take
$k$ strings of $\lceil \log_2 n \rceil$ bits from the source string and  reject and
resample those which specify the binary representation of a number greater
than $n-1$.  These $k$ strings, interpreted as the binary representation of $k$
numbers $i_1,\dots,i_k$, serve as the witnesses to test the primality of $n$ (i.e., the $i$ in $W(i,n)$).
Initially we take $k=1$ and increase $k$ until all the Carmichael numbers are
correctly determined to be composite.  

The metric for the test is taken to be the
smallest $k$ such that at most $k$ witness numbers were required to obtain a
verdict of non-primality for all of the Carmichael numbers.  
For each $k$, new bits are read from the sample string for each Carmichael number to be tested; we only restart reading from the start of the string (and thus recycling bits) when there was a
need to try a larger value of $k$ to pass this test. 

\begin{figure}[t]
	\begin{center}
		\includegraphics[width=0.6\textwidth]{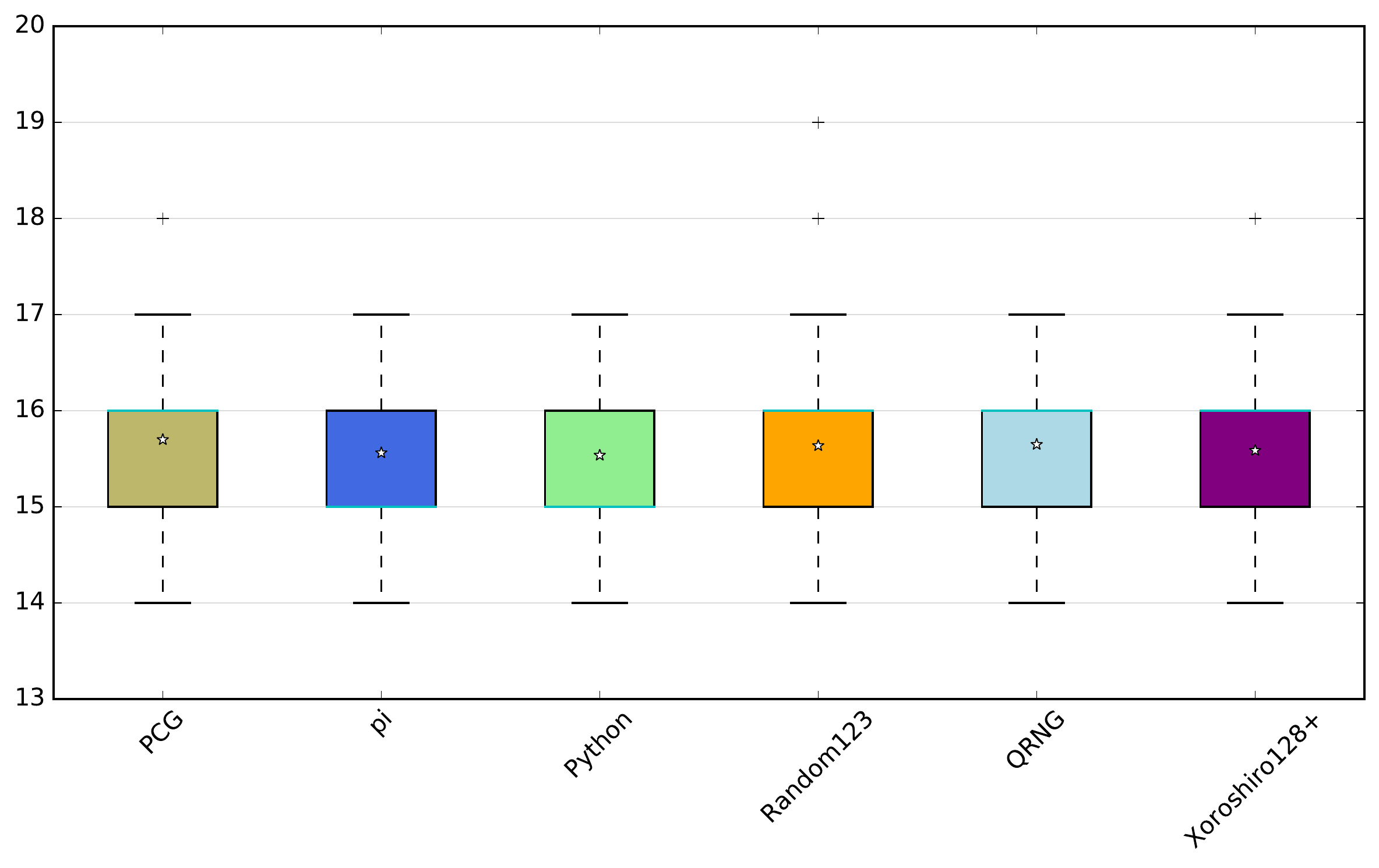}
		\caption{First Chaitin-Schwartz-Solovay-Strassen test on 80 samples: Box-plot showing the distribution in the minimum number 
		of witnesses needed to verify the compositeness of all Carmichael numbers of at most 16
		digits.}
		\label{box16mjd80}
	\end{center}
\end{figure}

Figure~\ref{box16mjd80} shows the performance of the 80 bit strings from each RNG (i.e., the same ones as tested for Borel normality in Section~\ref{sec:BN}) using the metric described above. 

The full results of the statistical analysis of this test (as well as the following) are given in the Appendix.
The Kolmogorov-Smirnov tests found no statistical significant difference between any of the sources of randomness (see Table~\ref{tab:CSSStest1CW16KS}).
The Shapiro-Wilk tests showed that the distribution of test statistics were not normally distributed (see Table~\ref{tab:CSSStest1CW16SW}), so further parametric tests were not performed.
This test therefore did not provide any evidence of significant differences between the RNGs, let alone evidence of incomputability of the QRNG.

\subsubsection{Second Chaitin-Schwartz-Solovay-Strassen test}

We next consider a closely related (and similarly motivated) test with a slightly different metric.
For each Carmichael number $n$, we repeatedly obtain a witness from the string being tested (in the same manner as in the first test and using new bits for each Carmichael number) until  the compositeness of $n$ is successfully witnessed. 
For this test metric we take the total number of bits used (for a given string to test) to confirm the compositeness of all 16 digit Carmichael numbers.
We calculate this as the sum, over all such Carmichael numbers $n$, of $\lceil \log_2 n \rceil$  times the number of
Solovay-Strassen trials needed to witness the compositeness of $n$.
(In this way, bits that are read but then rejected because they give a witness larger than $n$ do not contribute to the metric.)

\begin{figure}[ht]
	\begin{center}
		\begin{tabular}{cc}
		\includegraphics[width=0.48\textwidth]{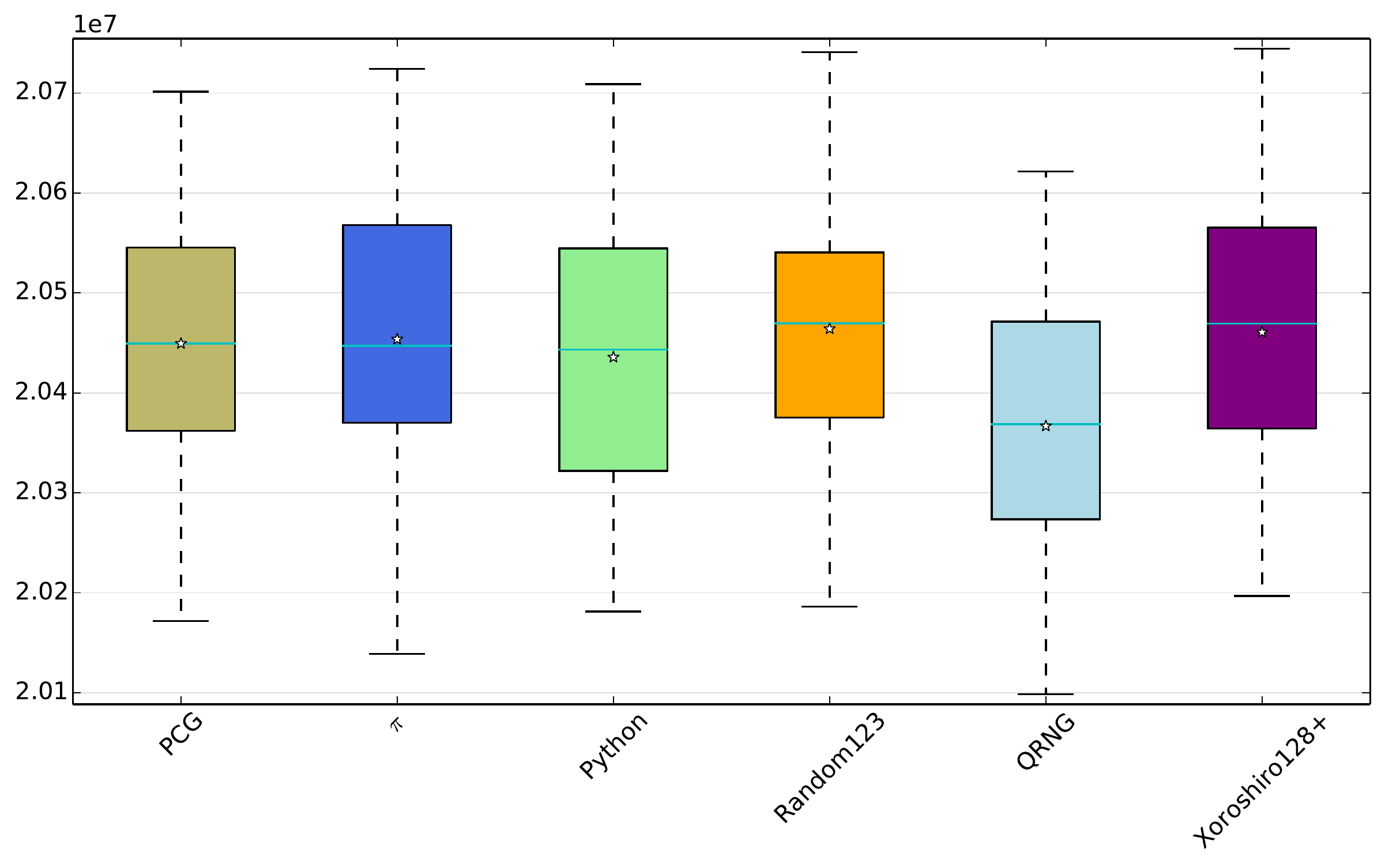} & \includegraphics[width=0.48\textwidth]{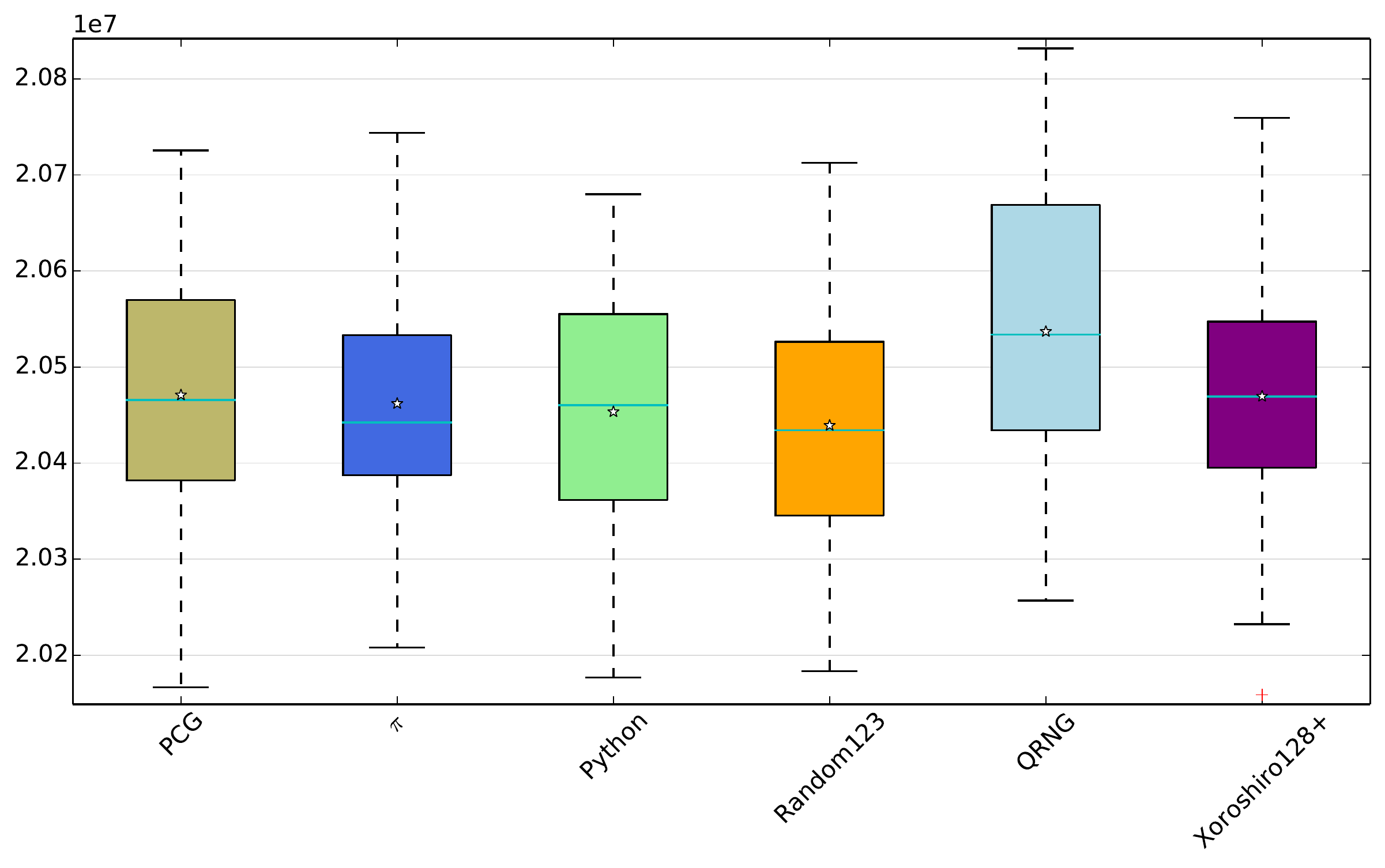}\\
		(a) & (b)
		\end{tabular}
		\caption{Second Chaitin-Schwartz-Solovay-Strassen test: total number of bits
		required to verify the compositeness of all Carmichael numbers of at most 16 digits using (a) the 80 strings from each RNG, and (b) the complement of these strings.
		}
		\label{boxcount16}
	\end{center}
\end{figure}

Figure~\ref{boxcount16}(a) shows a boxplot of the results for the 80 strings from each RNG being tested. 
The visible difference between the QRNG and the other sources is confirmed by the Kolmogorov-Smirnov tests (see Table~\ref{tab:CSSStest1CbOrKS}), which showed a statistically significant difference between the QRNG and $\pi$, Random123 and xoroshiro128+.
There is not, however, a general trend of normality for the test metric across all sources (in particular, there is weak evidence to reject normality of the distribution for the Python strings; see Table~\ref{tab:CSSStest1CbOrSW}), so it is not appropriate to use Welch's $t$-test to look for a difference between the QRNG and Python.

Although a significant difference was found between the QRNG and most the other sources, this is not necessarily a result of the incomputability we wish to test.
Indeed, we have already seen from the Borel normality test that the QRNG has a small statistical bias, so we should thus verify that the difference seen here is not also a result of this bias.
As mentioned earlier, we opted against trying to normalise the data to see if the bias is indeed to origin of the effect, not only because, with the amount of data available to us, this would markedly reduce our statistical power, but also because the effect of normalisation on the algorithmic complexity of the sequence is not entirely understood.
Instead, a simple way to test whether bias is behind the observed differences is to perform the same test on the complement of the strings we have tested (i.e., exchanging 0 and 1).
Since this transformation preserves randomness and incomputability, if the difference observed is evidence of such properties it should not be affected by such a transformation.

Figure~\ref{boxcount16}(b) shows the result of the test on the complemented sequences.
Here we see that again there is an apparent difference between the QRNG and some of the other sources.
This is confirmed by the Kolmogorov-Smirnov tests (see Table~\ref{tab:CSSStest1CbCoKS}) to be the case between the QRNG and $\pi$, Python and Random123.
In this case, the test metric is consistent with being normally distributed (see Table~\ref{tab:CSSStest1CbCoSW}), so it is reasonable to use Welch's $t$-test to try and confirm this difference further under an assumption of normality.
Doing so (see Table~\ref{tab:CSSStest1CbCoWt}) shows that there is indeed a statistically significant difference between the QRNG and all the other sources on the complemented strings.

However, as is clear from Figure~\ref{boxcount16}(b), this difference is in the opposite direction to (and of the same magnitude as) that in Figure~\ref{boxcount16}(a): in the latter the QRNG appears to perform better, while in the former, it performs worse.
It thus appears that this difference was indeed due to the bias of the QRNG rather than incomputability.
Nonetheless, we note that it is strange that biased sequences (in particular, biased towards having more zeroes) perform better in proving the compositeness of Carmichael numbers; we are not aware of any number theoretic explanation for this.

To conclude, this test shows that the QRNG behaves significantly
differently from almost all the other sources on this test (whether we use either the original bits or
the complemented bits), but that this difference is likely due  to the bias of the QRNG.
Understanding better why this bias makes such a difference would nonetheless be interesting.

\subsubsection{Third Chaitin-Schwartz-Solovay-Strassen test}
\label{sec:CSSStest3}

While the above tests are inspired by the Chaitin-Schwartz Theorem~\cite{Chaitin78}, they do not directly test the predicate $Z(s,n)$ appearing therein that we mentioned earlier. 
A key difference between these tests and the previous ones is the method they use to convert strings of random bits into potential witnesses to test.

Consider $s=s_{0} \dots s_{m-1}$ a binary string (of length $m$) and $n$ an
integer greater than 2.  Let $k$ be the smallest integer such that
$(n-1)^{k+1} >  2^m -1$; we can thus rewrite the number whose binary
representation is $s$ into base $n-1$ and obtain the unique string $d_k d_{k-1}
\dots d_0$ over the alphabet $\{0,1,\dots, n-2\}$, that is,   $$\sum_{i=0}^{k}
d_i(n-1)^{i}=\sum_{t=0}^{m-1}s_t 2^t.$$  The predicate $Z(s,n)$ is defined by
\begin{equation}\label{eq:Zpredicate}
	Z(s,J)= \neg W(1+d_0, n) \wedge \dots  \wedge  \neg W(1+d_{k-1}, n),
\end{equation}

where $W$ is the Solovay-Strassen predicate from Section~\ref{CSSStests}.
The digits of $s$ (rewritten in base $n-1$)  define the witnesses used to test the primality of $n$.

The main result from~\cite{Chaitin78} is:

\medskip

\begin{theorem}
	For all sufficiently large $c$, if  $s$ is  a $c$-Kolmogorov random string of length $\ell(\ell+2c)$ 
	and $n$ is an integer whose binary representation is $\ell$ bits long, then $Z(s,n)$ is true if and only if $n$ is prime.
\end{theorem}

In order to carry out these tests we first fix $c$.  For each Carmichael number
$n$ (with an $\ell$-bit binary representation) we  take
$c=\ell-1$.\footnote{This is somewhat arbitrary; other choices could of course
be made, but would make little difference to our test.}

\begin{figure}[ht]
	\begin{center}
		\begin{tabular}{cc}
		\includegraphics[width=0.48\textwidth]{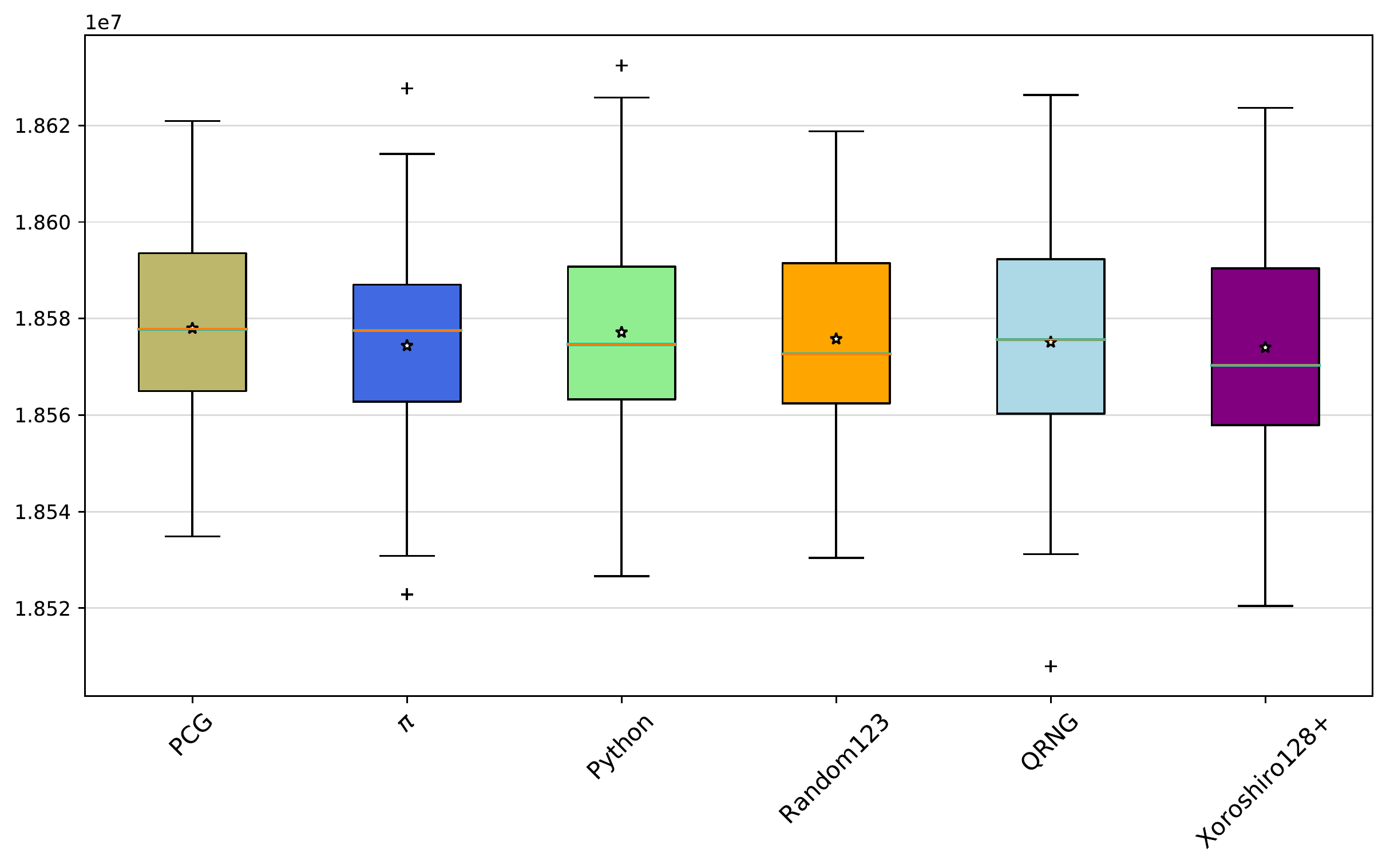} & \includegraphics[width=0.48\textwidth]{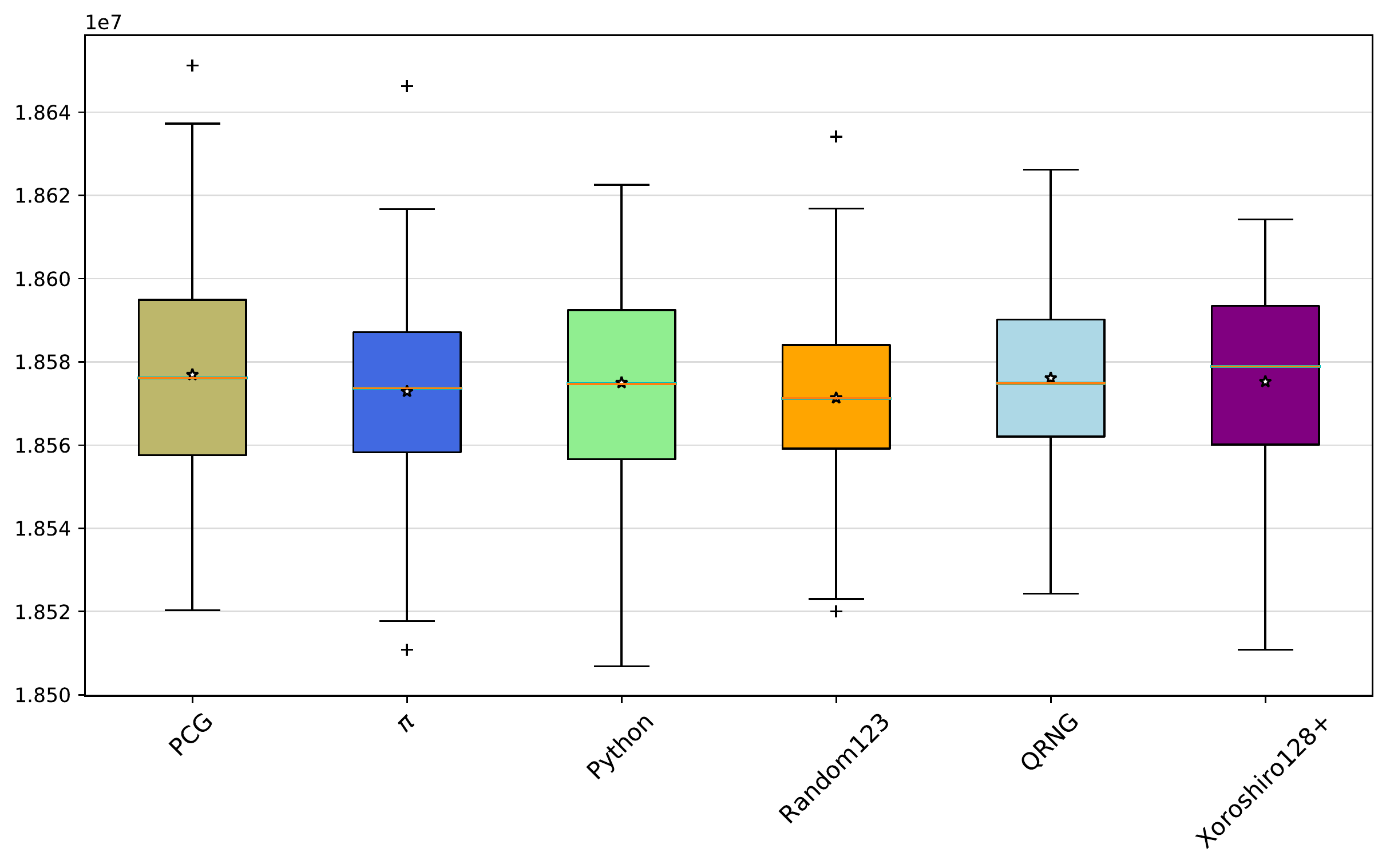} \\
		 (a) & (b)
		\end{tabular}
		
		\caption{Third Chaitin-Schwartz-Solovay-Strassen test: Box-plot showing the distribution of total number of bits used to identify all 16-digit Carmichael numbers as composite by (a) the 80 strings from each RNG, and (b) the complement of these strings.}		
		\label{boxplotCSSS2M1}
	\end{center}
\end{figure}

The metric of the third test has some similarities to that used in the second test.
For each such $n$ we take $\ell\,(\ell+2c)$ bits.
Rewriting $s$ in base $n-1$ as described above, we then compute $W(1+d_j,n)$ for $0\leq j \leq k$  until the
first $j$ is found such that $W(1+d_j,n)$ holds (and the compositeness of $n$
is thus witnessed).  The metric itself is then taken as the sum (over
all 16-digit Carmichael numbers $n$ tested) of $j\times \lceil\log_2(n-1)
\rceil $. Note that, if no first $j\le k$ is found such that $W(1+d_j,n)$ holds (which occurs very rarely), then we simply count 
all the bits used when testing that Carmichael number, i.e., $\ell\,(\ell+2c)$.
Figures~\ref{boxplotCSSS2M1}(a) shows
the performance of the 80 strings from each of the six sources according to
this metric.
In order to be able to do decouple any potential difference between the QRNG and the other sources due to algorithmic randomness from those resulting from the bias of the QRNG, we similarly perform the same test on the complement of each of the strings, the results of which are shown in Figure~\ref{boxplotCSSS2M1}(b).

The results of the Kolmogorov-Smirnov tests on the data shown in
Figures~\ref{boxplotCSSS2M1}(a) and~\ref{boxplotCSSS2M1}(b) are given in
Tables~\ref{tab:CSSS2M1_KS} and~\ref{tab:CSSS2M1C_KS}, respectively.
No statistically significant differences between any of the sources were found, reinforcing the impression given by Figure~\ref{boxplotCSSS2M1} that the RNGs all give similar results.
The Shapiro-Wilk test shows (see Tables~\ref{tab:CSSS2M1_SW} and~\ref{tab:CSSS2M1C_SW}) that there is no strong evidence against the normality of test metric for the non-complemented strings (but there was weak evidence against it for the complemented ones), so we were able to use Welch's $t$-test to look for any further evidence of differences between the sources on these strings (see Table~\ref{tab:CSSS2M1_Wt}).
No significant differences between the sources were found by these tests either.
We therefore conclude that
the third Chaitin-Schwartz-Solovay-Strassen test with this metric, which counts the
total number of bits required to verify the compositeness of all Carmichael
numbers of at most 16 digits, failed to find significant differences between the QRNG
and the PRNGs tested.

\subsubsection{Fourth Chaitin-Schwartz-Solovay-Strassen test}

The final test is based more closely on the Chaitin-Schwartz Theorem out of the tests we consider.
Rather than looking at how many witnesses need to be tested until a Carmichael number's compositeness is verified, we look directly at the ability of the entire \emph{set} of witnesses evaluated in~\eqref{eq:Zpredicate} to verify the compositeness of a number.
In other words, we look for direct violations of the Chaitin-Schwartz
Theorem:  a violation appears when for all $j=0,\dots,k-1$, $W(1+d_j,n)$ are false; that is,
all tests wrongly conclude that $n$ is ``probably prime''.

\begin{figure}[ht]
\begin{center}
	\begin{tabular}{cc}
	 \includegraphics[width=0.48\textwidth]{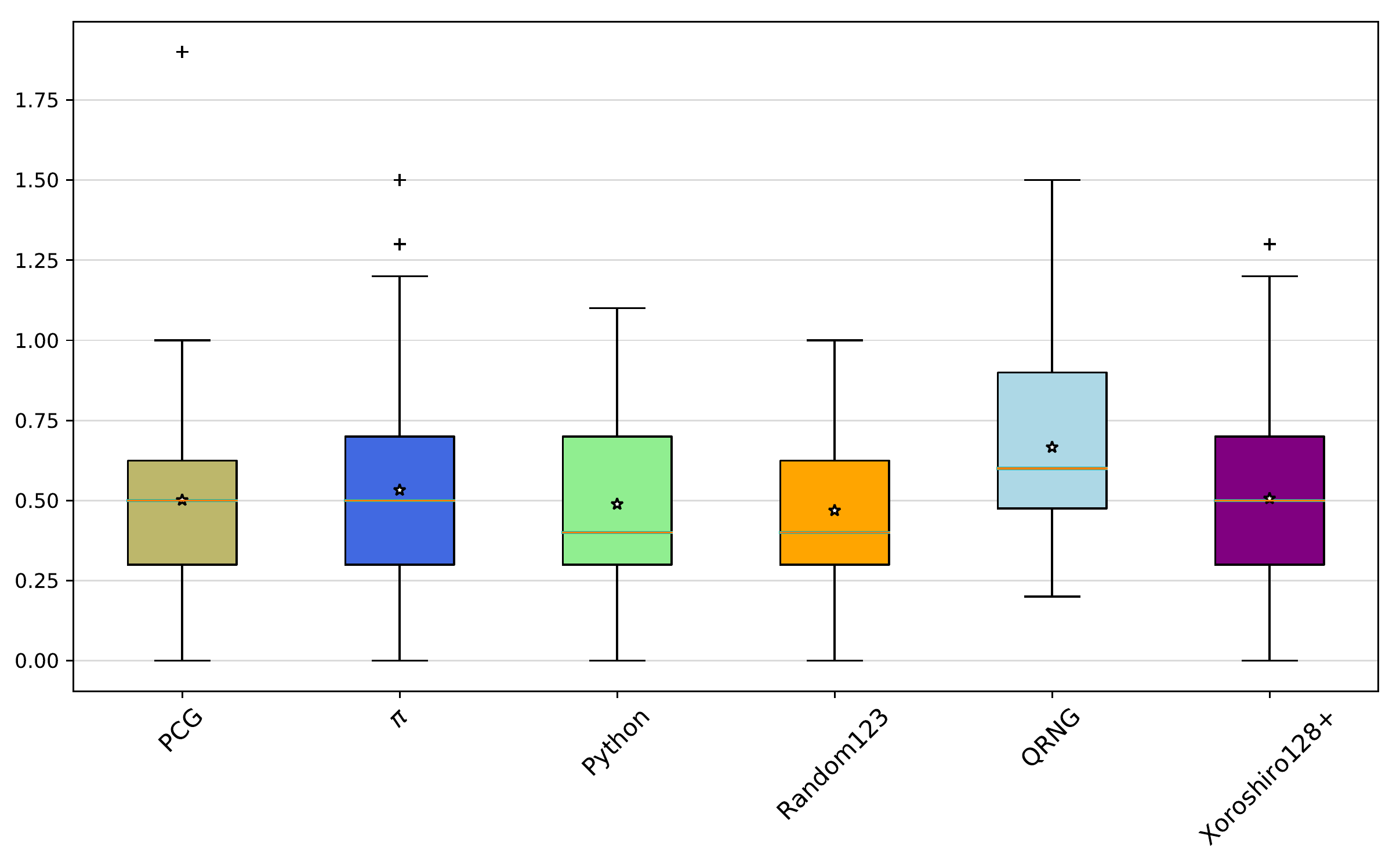} & \includegraphics[width=0.48\textwidth]{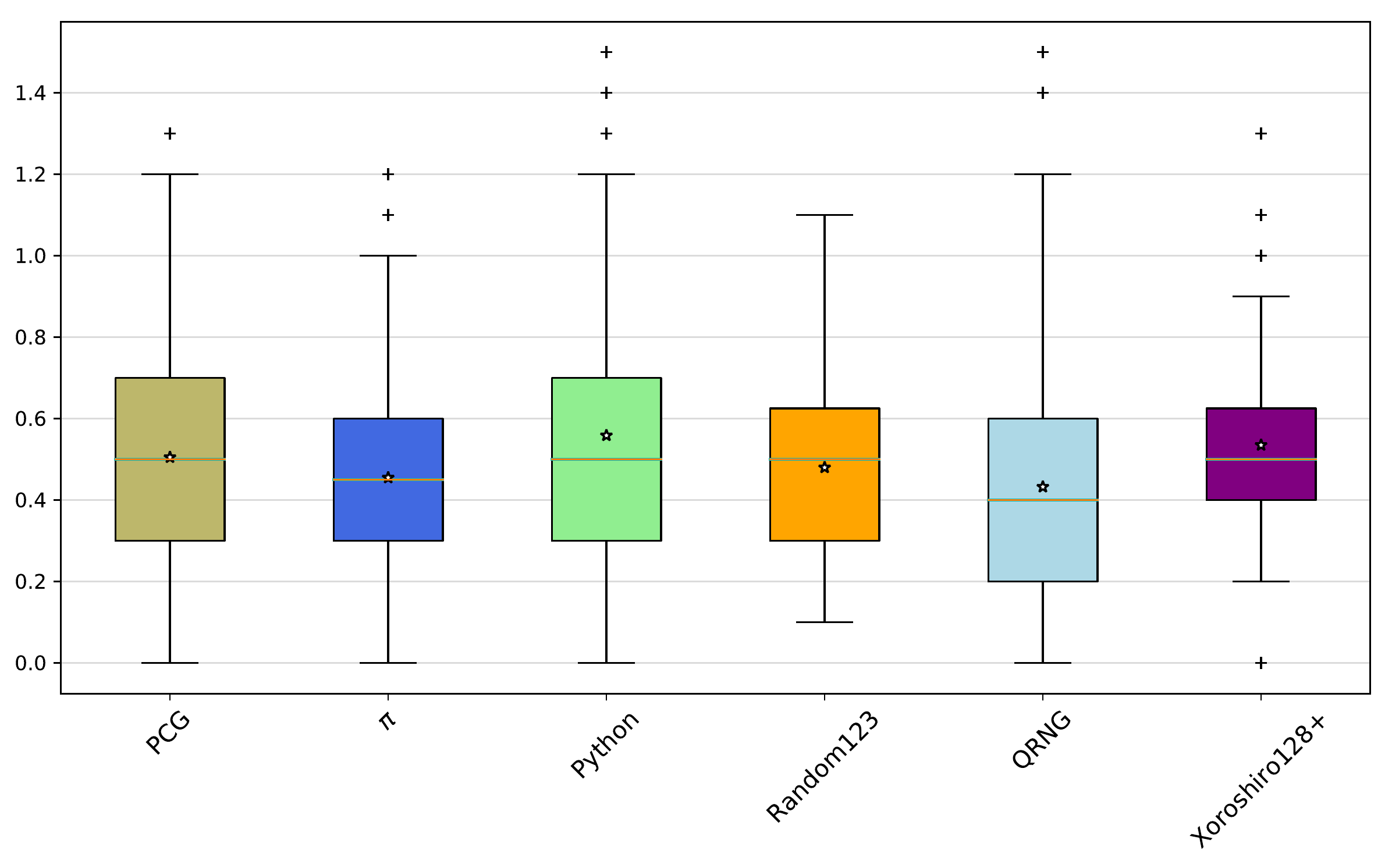}\\
	 (a) & (b)
	\end{tabular}
	\caption{Fourth Chaitin-Schwartz-Solovay-Strassen test: Box-plot showing the distribution of
		the average count of violations of the Chaitin-Schwartz Theorem for all odd composite numbers
		less than 50 by (a) the 80 strings from each RNG, and (b) the complement of these strings.}		
	\label{boxplotCSSS2M2}
	\end{center}
\end{figure}

However, as the Solovay-Strassen test guarantees that $W(1+d_j,n)$ is true with
probability at least one half when $n$ is a composite number,
it quickly becomes difficult, in practice, to observe such violations for even the smallest Carmichael numbers used in the previous tests.
In order to observe some violations
with the length of random strings (and time) we have access to, we have to severely
restrict ourselves and be content with testing the performance of the strings on only the odd composite numbers less than 50: $9,15,21,25,27,33,35,39,45,49$. 
For these numbers, we compute $Z(s,n)$ by reading $\ell(\ell+2c)$ bits and following the same procedure as in the third test. 
When $Z(s,n)=1 $, a violation of the Chaitin-Schwartz Theorem is thus observed. 
Since testing this predicate a single time on the ten numbers above would give insufficient statistics to observe any difference between the sources, we then repeated the above procedure reading from then 2nd bit of each string, then the 3rd, etc., until all the random bits have been used.
The metric is thereby taken as the average number of violations observed for the 10 composites tested (where the average is taken over all the repetitions). 
Figures~\ref{boxplotCSSS2M2}(a) and~\ref{boxplotCSSS2M2}(b) show  the results of this test for the 80
strings of each of the six sources used in the previous tests: again, the tests
in the former figure use the original strings from each source while the tests
in the latter use the complemented strings.

We apply the same statistical tests to determine whether there are any
statistically significant differences in performance between the different RNGs. 
The results of the Kolmogorov-Smirnov tests for the data in
Figures~\ref{boxplotCSSS2M2}(a) and~\ref{boxplotCSSS2M2}(b) are given in
Tables~\ref{tab:CSSS2M2_KS} and~\ref{tab:CSSS2M2C_KS}, respectively.  Unlike the results
for the previous metrics, the QRNG exhibits significantly different behaviour on the original (i.e., non-complemented) strings from the PCG, Python and Random123 PRNGs. However, no significant difference is found on any of the complemented strings. 
The Shapiro-Wilk tests (see Tables~\ref{tab:CSSS2M2_SW} and~\ref{tab:CSSS2M2C_SW}) find strong evidence against the normality of the distribution of the test metric, so Welch's $t$-test was not applied to see if further evidence of significant differences was present. 

Again, the reason for the apparently significant differences in performance between the QRNG and some of the sources (at least for the non-complemented strings) is unclear, and further investigation is required. 
The fact that only very small composite numbers were able to be tested means that, in the absence of strong evidence of differences between the sources, the results should be interpreted cautiously.
Indeed, the Chaitin-Schwartz Theorem is an asymptotic result, and  a significant difference on larger composites (ideally Carmichael numbers), would be preferable.
We thus cautiously conclude that the fourth Chaitin-Schwartz-Solovay-Strassen test with
the violation-count metric potentially identifies differences between QRNGs and the other sources, but that further testing and study is needed to confirm the robustness of the initial results observed here.

\section{Conclusions}

In this paper we looked at the ability to formulate tests that probed the incomputability and algorithmic randomness of strings produced by QRNGs.
Standard tests used to assess the quality of strings produced by PRNGs and QRNGs alike focus on simple statistical properties of the sequences, yet the most marked differences between QRNGs and PRNGs are the algorithmic properties of strings produced by such devices.
Such tests thus provide an important and novel approach to evaluating the performance of QRNGs.
This type of test, which probes the randomness of \emph{outputs} of QRNGs, is complementary to the certification of a QRNG as exploiting random \emph{processes}, either via theoretical analysis of the device~\cite{Abbott:2010fk,Abbott:2012fk} or the use of device-independent randomness expansion schemes~\cite{Pironio:2010aa,Colbeck:2011gd}.

The properties of incomputability (and, consequently, of algorithmic randomness too) mean that one must resort to indirect tests of incomputability in practice, and we discussed several such approaches.
We considered testing the Borel normality of sequences---a necessary property of algorithmic randomness---which probes the bias of a sequence rather than its incomputability \emph{per se}. This served as a useful preliminary probe for the analysis of later tests.
We then focused on a different approach based around the Chaitin-Schwartz Theorem, which shows a practical consequence of algorithmic randomness in probabilistic primality testing algorithms.
We proposed four different tests based on this result which, in principle, could exhibit advantages due to the incomputability---as well as the algorithmic randomness---of sequences from QRNGs over PRNGs.

To assess the practical utility of these tests, we applied them to long sequences generated by various RNGs: a QRNG (described in Section~\ref{sec:QRNG}), and several different PRNGs.
Two of the tests (the first and the third) failed to find any significant differences between the QRNG and the PRNGs.
A significant difference was, on the other hand, observed, for the second test. However, we were able to show that the difference was due to a small bias present in the strings produced by the QRNG rather than a result of any incomputability.
Indeed, this highlighted a key challenge: the need to decouple the incomputability from the bias within the test results, since the tests can in general be affected by both these elements.
To this end, we examined the performance of tests on the complement of the strings as well as the strings themselves, but conclude that care should be taken to formulate tests that are not affected by the bias of a sequence.
This task is complicated, however, by the fact that the effect of using a biased distribution in probabilistic primality testing is not well understood theoretically.
For future studies, it would thus be desirable to have sufficiently long strings to analyse from a certified QRNG for which the bias is sufficiently small so as to not be a limiting factor for the tests. Conversely, one should also study further the effect of normalisation procedures~\cite{Neumann:2012uq,Abbott:2010ij} on such metrics, so that such tests can be properly analysed when applied to normalised, rather than raw, sequences of bits.

Our fourth test, which was designed to follow more faithfully the Chaitin-Schwartz Theorem and to be potentially more robust to bias (but, unfortunately, more demanding to apply in practice), produced ambiguous results.
In particular, significant differences were found only on the non-complemented strings, but it was not clear whether these differences were entirely due to bias, as one would expect the complemented strings to show a similar difference in the opposite direction, which was not observed.
Due to the practical limitations of this test and small numbers tested, further testing (and, probably, refinements of the test itself) on more data are needed to understand this effect better.

While our tests failed to find any conclusive experimental evidence of incomputability of quantum randomness, they provide an important study for the development of new types of tests aimed at probing algorithmic properties of quantum randomness.
Indeed, being based on the Chaitin-Schwartz Theorem, the tests in fact probe the stronger property of $c$-Kolmogorov randomness, and this fact potentially contributes to the difficulty in observing indirect effects of incomputability.
The development of further tests to this end, as well as additional experimental studies, are therefore merited.

We finish by reiterating that tests of the output of QRNGs, such as we describe here, complement, rather than replace, the certification by value indefiniteness of a QRNG.
Indeed, just as with standard statistical tests of RNGs, even if no difference between QRNGs and PRNGs is found on the tests, it is important that QRNGs are verified to pass such tests.
QRNGs certified by the Kochen-Specker Theorem, such as the one used to provide the data for this paper~\cite{PhysRevLett.119.240501}, can also be used to perform device-independent randomness expansion~\cite{Um:2013tt,Deng13,Miller17}, and it would be interesting to combine algorithmic tests like we develop here with such an approach, although producing sufficient amounts of data for this to be possible remain a challenge.

Lastly, we note that all the test data (i.e., random strings), programs and results are available online in~\cite{CDMTCS515V2}. 

\subsection*{Acknowledgements}

The authors acknowledge fruitful discussions with Arkady Fedorov, Anatoly Kulikov, Frank Stephan and Karl Svozil.
A.\ A.\ Abbott acknowledges financial support from the `Retour Post-Doctorants' (ANR-13-PDOC-0026) programme of the French National Research Agency.

\bibliographystyle{apsrev4-1_modified}
\bibliography{experimental_randomness_analysis}

\begin{thebibliography}{75}%
\makeatletter
\providecommand \@ifxundefined [1]{%
 \@ifx{#1\undefined}
}%
\providecommand \@ifnum [1]{%
 \ifnum #1\expandafter \@firstoftwo
 \else \expandafter \@secondoftwo
 \fi
}%
\providecommand \@ifx [1]{%
 \ifx #1\expandafter \@firstoftwo
 \else \expandafter \@secondoftwo
 \fi
}%
\providecommand \natexlab [1]{#1}%
\providecommand \enquote  [1]{#1}%
\providecommand \bibnamefont  [1]{#1}%
\providecommand \bibfnamefont [1]{#1}%
\providecommand \citenamefont [1]{#1}%
\providecommand \href@noop [0]{\@secondoftwo}%
\providecommand \href [0]{\begingroup \@sanitize@url \@href}%
\providecommand \@href[1]{\@@startlink{#1}\@@href}%
\providecommand \@@href[1]{\endgroup#1\@@endlink}%
\providecommand \@sanitize@url [0]{\catcode `\\12\catcode `\$12\catcode
  `\&12\catcode `\#12\catcode `\^12\catcode `\_12\catcode `\%12\relax}%
\providecommand \@@startlink[1]{}%
\providecommand \@@endlink[0]{}%
\providecommand \url  [0]{\begingroup\@sanitize@url \@url }%
\providecommand \@url [1]{\endgroup\@href {#1}{\urlprefix }}%
\providecommand \urlprefix  [0]{URL }%
\providecommand \Eprint [0]{\href }%
\providecommand \doibase [0]{http://dx.doi.org/}%
\providecommand \selectlanguage [0]{\@gobble}%
\providecommand \bibinfo  [0]{\@secondoftwo}%
\providecommand \bibfield  [0]{\@secondoftwo}%
\providecommand \translation [1]{[#1]}%
\providecommand \BibitemOpen [0]{}%
\providecommand \bibitemStop [0]{}%
\providecommand \bibitemNoStop [0]{.\EOS\space}%
\providecommand \EOS [0]{\spacefactor3000\relax}%
\providecommand \BibitemShut  [1]{\csname bibitem#1\endcsname}%
\let\auto@bib@innerbib\@empty
\bibitem [{\citenamefont {Lenstra}\ \emph {et~al.}(2012)\citenamefont
  {Lenstra}, \citenamefont {Hughes}, \citenamefont {Augier}, \citenamefont
  {Bos}, \citenamefont {Kleinjung},\ and\ \citenamefont
  {Wachter}}]{factor_wrong2012}%
  \BibitemOpen
  \bibfield  {author} {\bibinfo {author} {\bibfnamefont {A.~K.}\ \bibnamefont
  {Lenstra}}, \bibinfo {author} {\bibfnamefont {J.~P.}\ \bibnamefont {Hughes}},
  \bibinfo {author} {\bibfnamefont {M.}~\bibnamefont {Augier}}, \bibinfo
  {author} {\bibfnamefont {J.~W.}\ \bibnamefont {Bos}}, \bibinfo {author}
  {\bibfnamefont {T.}~\bibnamefont {Kleinjung}}, \ and\ \bibinfo {author}
  {\bibfnamefont {C.}~\bibnamefont {Wachter}},\ }\href@noop {} {\enquote
  {\bibinfo {title} {Ron was wrong, {W}hit is right},}\ } (\bibinfo {year}
  {2012}),\ \bibinfo {note} {santa Barbara: IACR: 17,
  \url{https://eprint.iacr.org/2012/064.pdf}}\BibitemShut {NoStop}%
\bibitem [{\citenamefont {Svozil}(1990)}]{Svozil:1990aa}%
  \BibitemOpen
  \bibfield  {author} {\bibinfo {author} {\bibfnamefont {K.}~\bibnamefont
  {Svozil}},\ }\href {\doibase 10.1016/0375-9601(90)90408-G} {\bibfield
  {journal} {\bibinfo  {journal} {Physics Letters A}\ }\textbf {\bibinfo
  {volume} {143}},\ \bibinfo {pages} {433} (\bibinfo {year}
  {1990})}\BibitemShut {NoStop}%
\bibitem [{\citenamefont {Stefanov}\ \emph {et~al.}(2000)\citenamefont
  {Stefanov}, \citenamefont {Gisin}, \citenamefont {Guinnard}, \citenamefont
  {Guinnard},\ and\ \citenamefont {Zbinden}}]{Stefanov:2000aa}%
  \BibitemOpen
  \bibfield  {author} {\bibinfo {author} {\bibfnamefont {A.}~\bibnamefont
  {Stefanov}}, \bibinfo {author} {\bibfnamefont {N.}~\bibnamefont {Gisin}},
  \bibinfo {author} {\bibfnamefont {O.}~\bibnamefont {Guinnard}}, \bibinfo
  {author} {\bibfnamefont {L.}~\bibnamefont {Guinnard}}, \ and\ \bibinfo
  {author} {\bibfnamefont {H.}~\bibnamefont {Zbinden}},\ }\href {\doibase
  10.1080/09500340008233380} {\bibfield  {journal} {\bibinfo  {journal}
  {Journal of Modern Optics}\ }\textbf {\bibinfo {volume} {47}},\ \bibinfo
  {pages} {595} (\bibinfo {year} {2000})}\BibitemShut {NoStop}%
\bibitem [{\citenamefont {Pironio}\ \emph {et~al.}(2010)\citenamefont
  {Pironio}, \citenamefont {Ac{\'{i}}n}, \citenamefont {Massar}, \citenamefont
  {de~la Giroday}, \citenamefont {Matsukevich}, \citenamefont {Maunz},
  \citenamefont {Olmchenk}, \citenamefont {Hayes}, \citenamefont {Luo},
  \citenamefont {Manning},\ and\ \citenamefont {Monroe}}]{Pironio:2010aa}%
  \BibitemOpen
  \bibfield  {author} {\bibinfo {author} {\bibfnamefont {S.}~\bibnamefont
  {Pironio}}, \bibinfo {author} {\bibfnamefont {A.}~\bibnamefont {Ac{\'{i}}n}},
  \bibinfo {author} {\bibfnamefont {S.}~\bibnamefont {Massar}}, \bibinfo
  {author} {\bibfnamefont {A.~B.}\ \bibnamefont {de~la Giroday}}, \bibinfo
  {author} {\bibfnamefont {D.~N.}\ \bibnamefont {Matsukevich}}, \bibinfo
  {author} {\bibfnamefont {P.}~\bibnamefont {Maunz}}, \bibinfo {author}
  {\bibfnamefont {S.}~\bibnamefont {Olmchenk}}, \bibinfo {author}
  {\bibfnamefont {D.}~\bibnamefont {Hayes}}, \bibinfo {author} {\bibfnamefont
  {L.}~\bibnamefont {Luo}}, \bibinfo {author} {\bibfnamefont {T.~A.}\
  \bibnamefont {Manning}}, \ and\ \bibinfo {author} {\bibfnamefont
  {C.}~\bibnamefont {Monroe}},\ }\href {\doibase 10.1038/nature09008}
  {\bibfield  {journal} {\bibinfo  {journal} {Nature}\ }\textbf {\bibinfo
  {volume} {464}},\ \bibinfo {pages} {09008} (\bibinfo {year}
  {2010})}\BibitemShut {NoStop}%
\bibitem [{\citenamefont {Bera}\ \emph {et~al.}(2017)\citenamefont {Bera},
  \citenamefont {Ac{\'{i}}n}, \citenamefont {Ku\'{s}}, \citenamefont
  {Mitchell},\ and\ \citenamefont {Lewenstein}}]{Bera17}%
  \BibitemOpen
  \bibfield  {author} {\bibinfo {author} {\bibfnamefont {M.~N.}\ \bibnamefont
  {Bera}}, \bibinfo {author} {\bibfnamefont {A.}~\bibnamefont {Ac{\'{i}}n}},
  \bibinfo {author} {\bibfnamefont {M.}~\bibnamefont {Ku\'{s}}}, \bibinfo
  {author} {\bibfnamefont {M.}~\bibnamefont {Mitchell}}, \ and\ \bibinfo
  {author} {\bibfnamefont {M.}~\bibnamefont {Lewenstein}},\ }\href {\doibase
  10.1088/1361-6633/aa8731} {\bibfield  {journal} {\bibinfo  {journal} {Reports
  on Progress in Physics}\ }\textbf {\bibinfo {volume} {80}},\ \bibinfo {pages}
  {124001} (\bibinfo {year} {2017})}\BibitemShut {NoStop}%
\bibitem [{\citenamefont {Colbeck}\ and\ \citenamefont
  {Kent}(2011)}]{Colbeck:2011gd}%
  \BibitemOpen
  \bibfield  {author} {\bibinfo {author} {\bibfnamefont {R.}~\bibnamefont
  {Colbeck}}\ and\ \bibinfo {author} {\bibfnamefont {A.}~\bibnamefont {Kent}},\
  }\href {\doibase 10.1088/1751-8113/44/9/095305} {\bibfield  {journal}
  {\bibinfo  {journal} {Journal of Physics A: Mathematical and General}\
  }\textbf {\bibinfo {volume} {44}},\ \bibinfo {pages} {095305} (\bibinfo
  {year} {2011})}\BibitemShut {NoStop}%
\bibitem [{\citenamefont {Marsaglia}\ and\ \citenamefont
  {Zaman}(1990)}]{DIEHARD-paper}%
  \BibitemOpen
  \bibfield  {author} {\bibinfo {author} {\bibfnamefont {G.}~\bibnamefont
  {Marsaglia}}\ and\ \bibinfo {author} {\bibfnamefont {A.}~\bibnamefont
  {Zaman}},\ }\href {\doibase 10.1016/0167-7152(90)90092-L} {\bibfield
  {journal} {\bibinfo  {journal} {Statistics \& Probability Letters}\ }\textbf
  {\bibinfo {volume} {9}},\ \bibinfo {pages} {35} (\bibinfo {year} {1990})},\
  \bibinfo {note} {\url{http://www.stat.fsu.edu/pub/diehard/}}\BibitemShut
  {NoStop}%
\bibitem [{\citenamefont {Rukhin}\ \emph {et~al.}(2010)\citenamefont {Rukhin},
  \citenamefont {Soto}, \citenamefont {Nechvatal}, \citenamefont {Smid},
  \citenamefont {Barker}, \citenamefont {Leigh}, \citenamefont {Levenson},
  \citenamefont {Vangel}, \citenamefont {Banks}, \citenamefont {Heckert},
  \citenamefont {Dray},\ and\ \citenamefont {Vo}}]{NISTtests}%
  \BibitemOpen
  \bibfield  {author} {\bibinfo {author} {\bibfnamefont {A.}~\bibnamefont
  {Rukhin}}, \bibinfo {author} {\bibfnamefont {J.}~\bibnamefont {Soto}},
  \bibinfo {author} {\bibfnamefont {J.}~\bibnamefont {Nechvatal}}, \bibinfo
  {author} {\bibfnamefont {M.}~\bibnamefont {Smid}}, \bibinfo {author}
  {\bibfnamefont {E.}~\bibnamefont {Barker}}, \bibinfo {author} {\bibfnamefont
  {S.}~\bibnamefont {Leigh}}, \bibinfo {author} {\bibfnamefont
  {M.}~\bibnamefont {Levenson}}, \bibinfo {author} {\bibfnamefont
  {M.}~\bibnamefont {Vangel}}, \bibinfo {author} {\bibfnamefont
  {D.}~\bibnamefont {Banks}}, \bibinfo {author} {\bibfnamefont
  {A.}~\bibnamefont {Heckert}}, \bibinfo {author} {\bibfnamefont
  {J.}~\bibnamefont {Dray}}, \ and\ \bibinfo {author} {\bibfnamefont
  {S.}~\bibnamefont {Vo}},\ }\href@noop {} {\emph {\bibinfo {title} {A
  Statistical Test Suite for Random and Pseudorandom Number Generators for
  Cryptographic Applications}}},\ \bibinfo {type} {Special Publication}\
  \bibinfo {number} {800-22}\ (\bibinfo  {institution} {NIST},\ \bibinfo {year}
  {2010})\BibitemShut {NoStop}%
\bibitem [{\citenamefont {Abbott}\ \emph
  {et~al.}(2014{\natexlab{a}})\citenamefont {Abbott}, \citenamefont {Calude},\
  and\ \citenamefont {Svozil}}]{Abbott:2010fk}%
  \BibitemOpen
  \bibfield  {author} {\bibinfo {author} {\bibfnamefont {A.~A.}\ \bibnamefont
  {Abbott}}, \bibinfo {author} {\bibfnamefont {C.~S.}\ \bibnamefont {Calude}},
  \ and\ \bibinfo {author} {\bibfnamefont {K.}~\bibnamefont {Svozil}},\ }\href
  {\doibase 10.1017/S0960129512000692} {\bibfield  {journal} {\bibinfo
  {journal} {Mathematical Structures in Computer Science}\ }\textbf {\bibinfo
  {volume} {24}},\ \bibinfo {pages} {e240303} (\bibinfo {year}
  {2014}{\natexlab{a}})}\BibitemShut {NoStop}%
\bibitem [{\citenamefont {Abbott}\ \emph {et~al.}(2012)\citenamefont {Abbott},
  \citenamefont {Calude}, \citenamefont {Conder},\ and\ \citenamefont
  {Svozil}}]{Abbott:2012fk}%
  \BibitemOpen
  \bibfield  {author} {\bibinfo {author} {\bibfnamefont {A.~A.}\ \bibnamefont
  {Abbott}}, \bibinfo {author} {\bibfnamefont {C.~S.}\ \bibnamefont {Calude}},
  \bibinfo {author} {\bibfnamefont {J.}~\bibnamefont {Conder}}, \ and\ \bibinfo
  {author} {\bibfnamefont {K.}~\bibnamefont {Svozil}},\ }\href {\doibase
  10.1103/PhysRevA.86.062109} {\bibfield  {journal} {\bibinfo  {journal}
  {Physical Review A}\ }\textbf {\bibinfo {volume} {86}},\ \bibinfo {pages}
  {062109} (\bibinfo {year} {2012})}\BibitemShut {NoStop}%
\bibitem [{\citenamefont {Calude}\ and\ \citenamefont
  {Svozil}(2008)}]{Calude:2008aa}%
  \BibitemOpen
  \bibfield  {author} {\bibinfo {author} {\bibfnamefont {C.~S.}\ \bibnamefont
  {Calude}}\ and\ \bibinfo {author} {\bibfnamefont {K.}~\bibnamefont
  {Svozil}},\ }\href {\doibase 10.1166/asl.2008.016} {\bibfield  {journal}
  {\bibinfo  {journal} {Advanced Science Letters}\ }\textbf {\bibinfo {volume}
  {1}},\ \bibinfo {pages} {165} (\bibinfo {year} {2008})}\BibitemShut {NoStop}%
\bibitem [{\citenamefont {Abbott}\ \emph
  {et~al.}(2015{\natexlab{a}})\citenamefont {Abbott}, \citenamefont {Calude},\
  and\ \citenamefont {Svozil}}]{Abbott:2015fk}%
  \BibitemOpen
  \bibfield  {author} {\bibinfo {author} {\bibfnamefont {A.~A.}\ \bibnamefont
  {Abbott}}, \bibinfo {author} {\bibfnamefont {C.~S.}\ \bibnamefont {Calude}},
  \ and\ \bibinfo {author} {\bibfnamefont {K.}~\bibnamefont {Svozil}},\ }in\
  \href {\doibase 10.1007/978-3-319-23534-9_4} {\emph {\bibinfo {booktitle}
  {Fields of Logic and Computation {II} -- Essays Dedicated to Yuri Gurevich on
  the Occasion of His 75th Birthday}}},\ \bibinfo {series} {Lecture Notes in
  Computer Science}, Vol.\ \bibinfo {volume} {9300},\ \bibinfo {editor} {edited
  by\ \bibinfo {editor} {\bibfnamefont {L.~D.}\ \bibnamefont {Beklemishev}},
  \bibinfo {editor} {\bibfnamefont {A.}~\bibnamefont {Blass}}, \bibinfo
  {editor} {\bibfnamefont {N.}~\bibnamefont {Dershowitz}}, \bibinfo {editor}
  {\bibfnamefont {B.}~\bibnamefont {Finkbeiner}}, \ and\ \bibinfo {editor}
  {\bibfnamefont {W.}~\bibnamefont {Schulte}}}\ (\bibinfo  {publisher}
  {Springer International},\ \bibinfo {address} {Switzerland},\ \bibinfo {year}
  {2015})\ pp.\ \bibinfo {pages} {69--86}\BibitemShut {NoStop}%
\bibitem [{\citenamefont {Bar-Hillel}\ and\ \citenamefont
  {Wagenaar}(1991)}]{Bar-Hillel:1991aa}%
  \BibitemOpen
  \bibfield  {author} {\bibinfo {author} {\bibfnamefont {M.}~\bibnamefont
  {Bar-Hillel}}\ and\ \bibinfo {author} {\bibfnamefont {W.~A.}\ \bibnamefont
  {Wagenaar}},\ }\href {\doibase https://doi.org/10.1016/0196-8858(91)90029-I}
  {\bibfield  {journal} {\bibinfo  {journal} {Advances in Applied Mathematics}\
  }\textbf {\bibinfo {volume} {12}},\ \bibinfo {pages} {428} (\bibinfo {year}
  {1991})}\BibitemShut {NoStop}%
\bibitem [{\citenamefont {Figurska}\ \emph {et~al.}(2008)\citenamefont
  {Figurska}, \citenamefont {Sta\'{n}czyk},\ and\ \citenamefont
  {Kulesza}}]{consciouns_rand20018}%
  \BibitemOpen
  \bibfield  {author} {\bibinfo {author} {\bibfnamefont {M.}~\bibnamefont
  {Figurska}}, \bibinfo {author} {\bibfnamefont {M.}~\bibnamefont
  {Sta\'{n}czyk}}, \ and\ \bibinfo {author} {\bibfnamefont {K.}~\bibnamefont
  {Kulesza}},\ }\href {\doibase 10.1016/j.mehy.2007.06.038} {\bibfield
  {journal} {\bibinfo  {journal} {Medical Hypotheses}\ }\textbf {\bibinfo
  {volume} {70}},\ \bibinfo {pages} {182} (\bibinfo {year} {2008})}\BibitemShut
  {NoStop}%
\bibitem [{\citenamefont {Calude}(2002)}]{Calude:2002fk}%
  \BibitemOpen
  \bibfield  {author} {\bibinfo {author} {\bibfnamefont {C.~S.}\ \bibnamefont
  {Calude}},\ }\href@noop {} {\emph {\bibinfo {title} {Information and
  Randomness: An Algorithmic Perspective}}},\ \bibinfo {edition} {2nd}\ ed.\
  (\bibinfo  {publisher} {Springer-Verlag, Berlin},\ \bibinfo {year}
  {2002})\BibitemShut {NoStop}%
\bibitem [{\citenamefont {Borel}(1909)}]{Borel09}%
  \BibitemOpen
  \bibfield  {author} {\bibinfo {author} {\bibfnamefont {{\'E}.}~\bibnamefont
  {Borel}},\ }\href@noop {} {\bibfield  {journal} {\bibinfo  {journal}
  {Rendiconti del Circolo Matematico di Palermo}\ }\textbf {\bibinfo {volume}
  {27}},\ \bibinfo {pages} {247} (\bibinfo {year} {1909})}\BibitemShut
  {NoStop}%
\bibitem [{\citenamefont {Calude}(1994)}]{Calude:1994fk}%
  \BibitemOpen
  \bibfield  {author} {\bibinfo {author} {\bibfnamefont {C.~S.}\ \bibnamefont
  {Calude}},\ }in\ \href@noop {} {\emph {\bibinfo {booktitle} {Developments in
  Language Theory}}},\ \bibinfo {editor} {edited by\ \bibinfo {editor}
  {\bibfnamefont {G.}~\bibnamefont {Rozenberg}}\ and\ \bibinfo {editor}
  {\bibfnamefont {A.}~\bibnamefont {Salomaa}}}\ (\bibinfo  {publisher} {World
  Scientific, Singapore},\ \bibinfo {year} {1994})\ pp.\ \bibinfo {pages}
  {113--129}\BibitemShut {NoStop}%
\bibitem [{\citenamefont {Solovay}\ and\ \citenamefont
  {Strassen}(1977{\natexlab{a}})}]{Solovay77}%
  \BibitemOpen
  \bibfield  {author} {\bibinfo {author} {\bibfnamefont {R.}~\bibnamefont
  {Solovay}}\ and\ \bibinfo {author} {\bibfnamefont {V.}~\bibnamefont
  {Strassen}},\ }\href {\doibase 10.1137/0206006} {\bibfield  {journal}
  {\bibinfo  {journal} {SIAM Journal on Computing}\ }\textbf {\bibinfo {volume}
  {6}},\ \bibinfo {pages} {84} (\bibinfo {year} {1977}{\natexlab{a}})},\
  \bibinfo {note} {corrigendum in Ref.~\cite{Solovay77b}}\BibitemShut {NoStop}%
\bibitem [{\citenamefont {Calude}\ \emph {et~al.}(2010)\citenamefont {Calude},
  \citenamefont {Dinneen}, \citenamefont {Dumitrescu},\ and\ \citenamefont
  {Svozil}}]{Calude:2010aa}%
  \BibitemOpen
  \bibfield  {author} {\bibinfo {author} {\bibfnamefont {C.~S.}\ \bibnamefont
  {Calude}}, \bibinfo {author} {\bibfnamefont {M.~J.}\ \bibnamefont {Dinneen}},
  \bibinfo {author} {\bibfnamefont {M.}~\bibnamefont {Dumitrescu}}, \ and\
  \bibinfo {author} {\bibfnamefont {K.}~\bibnamefont {Svozil}},\ }\href
  {\doibase 10.1103/PhysRevA.82.022102} {\bibfield  {journal} {\bibinfo
  {journal} {Physical Review A}\ }\textbf {\bibinfo {volume} {82}},\ \bibinfo
  {pages} {022102} (\bibinfo {year} {2010})}\BibitemShut {NoStop}%
\bibitem [{\citenamefont {Chaitin}\ and\ \citenamefont
  {Schwartz}(1978)}]{Chaitin78}%
  \BibitemOpen
  \bibfield  {author} {\bibinfo {author} {\bibfnamefont {G.~J.}\ \bibnamefont
  {Chaitin}}\ and\ \bibinfo {author} {\bibfnamefont {J.~T.}\ \bibnamefont
  {Schwartz}},\ }\href {\doibase 10.1002/cpa.3160310407} {\bibfield  {journal}
  {\bibinfo  {journal} {Communications on Pure and Applied Mathematics}\
  }\textbf {\bibinfo {volume} {31}},\ \bibinfo {pages} {521} (\bibinfo {year}
  {1978})}\BibitemShut {NoStop}%
\bibitem [{\citenamefont {Kulikov}\ \emph {et~al.}(2017)\citenamefont
  {Kulikov}, \citenamefont {Jerger}, \citenamefont {Poto\ifmmode~\check{c}\else
  \v{c}\fi{}nik}, \citenamefont {Wallraff},\ and\ \citenamefont
  {Fedorov}}]{PhysRevLett.119.240501}%
  \BibitemOpen
  \bibfield  {author} {\bibinfo {author} {\bibfnamefont {A.}~\bibnamefont
  {Kulikov}}, \bibinfo {author} {\bibfnamefont {M.}~\bibnamefont {Jerger}},
  \bibinfo {author} {\bibfnamefont {A.}~\bibnamefont
  {Poto\ifmmode~\check{c}\else \v{c}\fi{}nik}}, \bibinfo {author}
  {\bibfnamefont {A.}~\bibnamefont {Wallraff}}, \ and\ \bibinfo {author}
  {\bibfnamefont {A.}~\bibnamefont {Fedorov}},\ }\href {\doibase
  10.1103/PhysRevLett.119.240501} {\bibfield  {journal} {\bibinfo  {journal}
  {Physical Review Letters}\ }\textbf {\bibinfo {volume} {119}},\ \bibinfo
  {pages} {240501} (\bibinfo {year} {2017})}\BibitemShut {NoStop}%
\bibitem [{\citenamefont {Champernowne}(1933)}]{Champernowne:1933kx}%
  \BibitemOpen
  \bibfield  {author} {\bibinfo {author} {\bibfnamefont {D.~G.}\ \bibnamefont
  {Champernowne}},\ }\href@noop {} {\bibfield  {journal} {\bibinfo  {journal}
  {Journal of the London Mathematical Society}\ }\textbf {\bibinfo {volume}
  {8}},\ \bibinfo {pages} {254} (\bibinfo {year} {1933})}\BibitemShut {NoStop}%
\bibitem [{\citenamefont {Chaitin}(1977)}]{Chaitin:1977fk}%
  \BibitemOpen
  \bibfield  {author} {\bibinfo {author} {\bibfnamefont {G.~J.}\ \bibnamefont
  {Chaitin}},\ }\href {\doibase 10.1147/rd.214.0350} {\bibfield  {journal}
  {\bibinfo  {journal} {IBM Journal of Research and Development}\ }\textbf
  {\bibinfo {volume} {21}},\ \bibinfo {pages} {350} (\bibinfo {year}
  {1977})}\BibitemShut {NoStop}%
\bibitem [{\citenamefont {Downey}\ and\ \citenamefont
  {Hirschfeldt}(2010)}]{DH}%
  \BibitemOpen
  \bibfield  {author} {\bibinfo {author} {\bibfnamefont {R.}~\bibnamefont
  {Downey}}\ and\ \bibinfo {author} {\bibfnamefont {D.}~\bibnamefont
  {Hirschfeldt}},\ }\href@noop {} {\emph {\bibinfo {title} {Algorithmic
  Randomness and Complexity}}}\ (\bibinfo  {publisher} {Springer},\ \bibinfo
  {address} {Berlin},\ \bibinfo {year} {2010})\BibitemShut {NoStop}%
\bibitem [{\citenamefont {Martin-L{\"{o}}f}(1966)}]{Martin-Lof:1966kx}%
  \BibitemOpen
  \bibfield  {author} {\bibinfo {author} {\bibfnamefont {P.}~\bibnamefont
  {Martin-L{\"{o}}f}},\ }\href {\doibase 10.1016/S0019-9958(66)80018-9}
  {\bibfield  {journal} {\bibinfo  {journal} {Information and Control}\
  }\textbf {\bibinfo {volume} {9}},\ \bibinfo {pages} {602} (\bibinfo {year}
  {1966})}\BibitemShut {NoStop}%
\bibitem [{\citenamefont {Calude}(2017)}]{Calude17}%
  \BibitemOpen
  \bibfield  {author} {\bibinfo {author} {\bibfnamefont {C.~S.}\ \bibnamefont
  {Calude}},\ }in\ \href {\doibase 10.1007/978-3-319-43669-2_11} {\emph
  {\bibinfo {booktitle} {The Incomputable: Journeys Beyond the {T}uring
  Barrier}}},\ \bibinfo {editor} {edited by\ \bibinfo {editor} {\bibfnamefont
  {S.~B.}\ \bibnamefont {Cooper}}\ and\ \bibinfo {editor} {\bibfnamefont
  {M.}~\bibnamefont {Soskova}}}\ (\bibinfo  {publisher} {Springer},\ \bibinfo
  {year} {2017})\ pp.\ \bibinfo {pages} {169--181}\BibitemShut {NoStop}%
\bibitem [{\citenamefont {Graham}\ and\ \citenamefont
  {Spencer}(1990)}]{Graham:1990oe}%
  \BibitemOpen
  \bibfield  {author} {\bibinfo {author} {\bibfnamefont {R.}~\bibnamefont
  {Graham}}\ and\ \bibinfo {author} {\bibfnamefont {J.~H.}\ \bibnamefont
  {Spencer}},\ }\href {\doibase 10.2307/2275058} {\bibfield  {journal}
  {\bibinfo  {journal} {Scientific American}\ }\textbf {\bibinfo {volume}
  {262}},\ \bibinfo {pages} {112} (\bibinfo {year} {1990})}\BibitemShut
  {NoStop}%
\bibitem [{\citenamefont {Abbott}(2015)}]{Abbott15}%
  \BibitemOpen
  \bibfield  {author} {\bibinfo {author} {\bibfnamefont {A.~A.}\ \bibnamefont
  {Abbott}},\ }\emph {\bibinfo {title} {Value Indefiniteness, Randomness and
  Unpredictability in Quantum Foundations}},\ \href@noop {} {Ph.D. thesis},\
  \bibinfo  {school} {University of Auckland; {\'E}cole Normale Sup{\'e}rieure
  de Paris} (\bibinfo {year} {2015})\BibitemShut {NoStop}%
\bibitem [{\citenamefont {von Neumann}(1963)}]{Neumann:2012uq}%
  \BibitemOpen
  \bibfield  {author} {\bibinfo {author} {\bibfnamefont {J.}~\bibnamefont {von
  Neumann}},\ }in\ \href@noop {} {\emph {\bibinfo {booktitle} {John von
  Neumann, Collected Works}}},\ \bibinfo {editor} {edited by\ \bibinfo {editor}
  {\bibfnamefont {A.~H.}\ \bibnamefont {Traub}}}\ (\bibinfo  {publisher}
  {MacMillan, New York},\ \bibinfo {year} {1963})\ pp.\ \bibinfo {pages}
  {768--770}\BibitemShut {NoStop}%
\bibitem [{\citenamefont {Eagle}(2014)}]{Eagle:2014jx}%
  \BibitemOpen
  \bibfield  {author} {\bibinfo {author} {\bibfnamefont {A.}~\bibnamefont
  {Eagle}},\ }in\ \href@noop {} {\emph {\bibinfo {booktitle} {The Stanford
  Encyclopedia of Philosophy}}},\ \bibinfo {editor} {edited by\ \bibinfo
  {editor} {\bibfnamefont {E.~N.}\ \bibnamefont {Zalta}}}\ (\bibinfo
  {publisher} {Stanford University},\ \bibinfo {year} {2014})\ \bibinfo
  {edition} {{Spring} 2014}\ ed.\BibitemShut {Stop}%
\bibitem [{\citenamefont {Solis}\ and\ \citenamefont
  {Hirsch}(2015)}]{Solis15b}%
  \BibitemOpen
  \bibfield  {author} {\bibinfo {author} {\bibfnamefont {A.}~\bibnamefont
  {Solis}}\ and\ \bibinfo {author} {\bibfnamefont {J.~G.}\ \bibnamefont
  {Hirsch}},\ }\href {\doibase 10.1088/1742-6596/624/1/012001} {\bibfield
  {journal} {\bibinfo  {journal} {Journal of Physics: Conference Series}\
  }\textbf {\bibinfo {volume} {624}},\ \bibinfo {pages} {012001} (\bibinfo
  {year} {2015})}\BibitemShut {NoStop}%
\bibitem [{\citenamefont {Eagle}(2005)}]{Eagle:2005ys}%
  \BibitemOpen
  \bibfield  {author} {\bibinfo {author} {\bibfnamefont {A.}~\bibnamefont
  {Eagle}},\ }\href {\doibase 10.1093/bjps/axi138} {\bibfield  {journal}
  {\bibinfo  {journal} {The British Journal for the Philosophy of Science}\
  }\textbf {\bibinfo {volume} {56}},\ \bibinfo {pages} {749} (\bibinfo {year}
  {2005})}\BibitemShut {NoStop}%
\bibitem [{\citenamefont {Abbott}\ \emph
  {et~al.}(2015{\natexlab{b}})\citenamefont {Abbott}, \citenamefont {Calude},\
  and\ \citenamefont {Svozil}}]{Abbott:2015vg}%
  \BibitemOpen
  \bibfield  {author} {\bibinfo {author} {\bibfnamefont {A.~A.}\ \bibnamefont
  {Abbott}}, \bibinfo {author} {\bibfnamefont {C.~S.}\ \bibnamefont {Calude}},
  \ and\ \bibinfo {author} {\bibfnamefont {K.}~\bibnamefont {Svozil}},\ }\href
  {\doibase 10.3390/info6040773} {\bibfield  {journal} {\bibinfo  {journal}
  {Information}\ }\textbf {\bibinfo {volume} {6}},\ \bibinfo {pages} {773}
  (\bibinfo {year} {2015}{\natexlab{b}})}\BibitemShut {NoStop}%
\bibitem [{\citenamefont {H{\'{a}}jek}(2014)}]{Hajek:2014di}%
  \BibitemOpen
  \bibfield  {author} {\bibinfo {author} {\bibfnamefont {A.}~\bibnamefont
  {H{\'{a}}jek}},\ }in\ \href@noop {} {\emph {\bibinfo {booktitle} {The
  Stanford Encyclopedia of Philosophy}}},\ \bibinfo {editor} {edited by\
  \bibinfo {editor} {\bibfnamefont {E.~N.}\ \bibnamefont {Zalta}}}\ (\bibinfo
  {publisher} {Stanford University},\ \bibinfo {year} {2014})\ \bibinfo
  {edition} {{Winter} 2012}\ ed.\BibitemShut {Stop}%
\bibitem [{\citenamefont {Gentle}(2003)}]{Gentle03}%
  \BibitemOpen
  \bibfield  {author} {\bibinfo {author} {\bibfnamefont {J.~E.}\ \bibnamefont
  {Gentle}},\ }\href {\doibase 10.1007/b97336} {\emph {\bibinfo {title} {Random
  Number Generations and {M}onte {C}arlo Methods}}},\ \bibinfo {edition} {2nd}\
  ed.\ (\bibinfo  {publisher} {Springer-Verlag},\ \bibinfo {address} {New
  York},\ \bibinfo {year} {2003})\BibitemShut {NoStop}%
\bibitem [{\citenamefont {Goldreich}(2001)}]{Goldreich01}%
  \BibitemOpen
  \bibfield  {author} {\bibinfo {author} {\bibfnamefont {O.}~\bibnamefont
  {Goldreich}},\ }\href@noop {} {\emph {\bibinfo {title} {Foundations of
  cryptography I: Basic Tools}}}\ (\bibinfo  {publisher} {Cambridge University
  Press},\ \bibinfo {address} {Cambridge},\ \bibinfo {year} {2001})\BibitemShut
  {NoStop}%
\bibitem [{\citenamefont {Bernstein}\ \emph {et~al.}(2013)\citenamefont
  {Bernstein}, \citenamefont {Chang}, \citenamefont {Cheng}, \citenamefont
  {Chou}, \citenamefont {Heninger}, \citenamefont {Lange},\ and\ \citenamefont
  {van Someren}}]{Bernstein13}%
  \BibitemOpen
  \bibfield  {author} {\bibinfo {author} {\bibfnamefont {D.~J.}\ \bibnamefont
  {Bernstein}}, \bibinfo {author} {\bibfnamefont {Y.-A.}\ \bibnamefont
  {Chang}}, \bibinfo {author} {\bibfnamefont {C.-M.}\ \bibnamefont {Cheng}},
  \bibinfo {author} {\bibfnamefont {L.-P.}\ \bibnamefont {Chou}}, \bibinfo
  {author} {\bibfnamefont {N.}~\bibnamefont {Heninger}}, \bibinfo {author}
  {\bibfnamefont {T.}~\bibnamefont {Lange}}, \ and\ \bibinfo {author}
  {\bibfnamefont {N.}~\bibnamefont {van Someren}},\ }in\ \href {\doibase
  10.1007/978-3-642-42045-0_18} {\emph {\bibinfo {booktitle} {Advances in
  Cryptology -- ASIACRYPT 2013}}},\ \bibinfo {editor} {edited by\ \bibinfo
  {editor} {\bibfnamefont {K.}~\bibnamefont {Sako}}\ and\ \bibinfo {editor}
  {\bibfnamefont {P.}~\bibnamefont {Sarkar}}}\ (\bibinfo  {publisher}
  {Springer},\ \bibinfo {address} {Berling},\ \bibinfo {year} {2013})\ pp.\
  \bibinfo {pages} {341--360}\BibitemShut {NoStop}%
\bibitem [{\citenamefont {Ac{\'{i}}n}(2013)}]{Acin:2013qa}%
  \BibitemOpen
  \bibfield  {author} {\bibinfo {author} {\bibfnamefont {A.}~\bibnamefont
  {Ac{\'{i}}n}},\ }in\ \href {\doibase 10.1007/978-1-4614-5212-6_2} {\emph
  {\bibinfo {booktitle} {Is Science Compatible with Free Will?: Exploring Free
  Will and Consciousness in the Light of Quantum Physics and Neuroscience}}},\
  \bibinfo {editor} {edited by\ \bibinfo {editor} {\bibfnamefont
  {A.}~\bibnamefont {Suarez}}\ and\ \bibinfo {editor} {\bibfnamefont
  {P.}~\bibnamefont {Adams}}}\ (\bibinfo  {publisher} {Springer-Verlag},\
  \bibinfo {address} {New York},\ \bibinfo {year} {2013})\ Chap.~\bibinfo
  {chapter} {2}, pp.\ \bibinfo {pages} {7--22}\BibitemShut {NoStop}%
\bibitem [{\citenamefont {Longo}\ and\ \citenamefont
  {Paul}(2008)}]{Longo:2008ud}%
  \BibitemOpen
  \bibfield  {author} {\bibinfo {author} {\bibfnamefont {G.}~\bibnamefont
  {Longo}}\ and\ \bibinfo {author} {\bibfnamefont {T.}~\bibnamefont {Paul}},\
  }in\ \href {\doibase 10.1142/9781848162778_0007} {\emph {\bibinfo {booktitle}
  {Computability in Context: Computation and Logic in the Real World}}},\
  \bibinfo {editor} {edited by\ \bibinfo {editor} {\bibfnamefont {S.~B.}\
  \bibnamefont {Cooper}}\ and\ \bibinfo {editor} {\bibfnamefont
  {A.}~\bibnamefont {Sorbi}}}\ (\bibinfo  {publisher} {Imperial College
  Press/World Scientific},\ \bibinfo {address} {London},\ \bibinfo {year}
  {2008})\ Chap.~\bibinfo {chapter} {7}, pp.\ \bibinfo {pages}
  {243--274}\BibitemShut {NoStop}%
\bibitem [{\citenamefont {Bell}(1964)}]{Bell:1964fk}%
  \BibitemOpen
  \bibfield  {author} {\bibinfo {author} {\bibfnamefont {J.~S.}\ \bibnamefont
  {Bell}},\ }\href {\doibase 10.1142/9789812386540_0002} {\bibfield  {journal}
  {\bibinfo  {journal} {Physics}\ }\textbf {\bibinfo {volume} {1}},\ \bibinfo
  {pages} {195} (\bibinfo {year} {1964})}\BibitemShut {NoStop}%
\bibitem [{\citenamefont {Aspect}\ \emph {et~al.}(1982)\citenamefont {Aspect},
  \citenamefont {Grangier},\ and\ \citenamefont {Roger}}]{Aspect:1982dp}%
  \BibitemOpen
  \bibfield  {author} {\bibinfo {author} {\bibfnamefont {A.}~\bibnamefont
  {Aspect}}, \bibinfo {author} {\bibfnamefont {P.}~\bibnamefont {Grangier}}, \
  and\ \bibinfo {author} {\bibfnamefont {G.}~\bibnamefont {Roger}},\ }\href
  {\doibase 10.1103/PhysRevLett.49.91} {\bibfield  {journal} {\bibinfo
  {journal} {Physical Review Letters}\ }\textbf {\bibinfo {volume} {49}},\
  \bibinfo {pages} {91} (\bibinfo {year} {1982})}\BibitemShut {NoStop}%
\bibitem [{\citenamefont {Kochen}\ and\ \citenamefont
  {Specker}(1967)}]{Kochen:1967fk}%
  \BibitemOpen
  \bibfield  {author} {\bibinfo {author} {\bibfnamefont {S.~B.}\ \bibnamefont
  {Kochen}}\ and\ \bibinfo {author} {\bibfnamefont {E.}~\bibnamefont
  {Specker}},\ }\href {\doibase 10.1512/iumj.1968.17.17004} {\bibfield
  {journal} {\bibinfo  {journal} {Journal of Mathematics and Mechanics (now
  Indiana University Mathematics Journal)}\ }\textbf {\bibinfo {volume} {17}},\
  \bibinfo {pages} {59} (\bibinfo {year} {1967})}\BibitemShut {NoStop}%
\bibitem [{\citenamefont {Abbott}\ \emph {et~al.}(2013)\citenamefont {Abbott},
  \citenamefont {Calude},\ and\ \citenamefont {Svozil}}]{Abbott:2013ly}%
  \BibitemOpen
  \bibfield  {author} {\bibinfo {author} {\bibfnamefont {A.~A.}\ \bibnamefont
  {Abbott}}, \bibinfo {author} {\bibfnamefont {C.~S.}\ \bibnamefont {Calude}},
  \ and\ \bibinfo {author} {\bibfnamefont {K.}~\bibnamefont {Svozil}},\ }\href
  {\doibase 10.1103/PhysRevA.89.032109} {\bibfield  {journal} {\bibinfo
  {journal} {Physical Review A}\ }\textbf {\bibinfo {volume} {89}},\ \bibinfo
  {pages} {032109} (\bibinfo {year} {2013})}\BibitemShut {NoStop}%
\bibitem [{\citenamefont {Abbott}\ \emph
  {et~al.}(2015{\natexlab{c}})\citenamefont {Abbott}, \citenamefont {Calude},\
  and\ \citenamefont {Svozil}}]{Abbott:2015bx}%
  \BibitemOpen
  \bibfield  {author} {\bibinfo {author} {\bibfnamefont {A.~A.}\ \bibnamefont
  {Abbott}}, \bibinfo {author} {\bibfnamefont {C.~S.}\ \bibnamefont {Calude}},
  \ and\ \bibinfo {author} {\bibfnamefont {K.}~\bibnamefont {Svozil}},\ }\href
  {\doibase 10.1063/1.4931658} {\bibfield  {journal} {\bibinfo  {journal}
  {Journal of Mathematical Physics}\ }\textbf {\bibinfo {volume} {56}},\
  \bibinfo {pages} {102201} (\bibinfo {year} {2015}{\natexlab{c}})}\BibitemShut
  {NoStop}%
\bibitem [{\citenamefont {Schmidt}(1970)}]{Schmidt:1970aa}%
  \BibitemOpen
  \bibfield  {author} {\bibinfo {author} {\bibfnamefont {H.}~\bibnamefont
  {Schmidt}},\ }\href {\doibase 10.1063/1.1658698} {\bibfield  {journal}
  {\bibinfo  {journal} {Journal of Applied Physics}\ }\textbf {\bibinfo
  {volume} {41}},\ \bibinfo {pages} {462} (\bibinfo {year} {1970})}\BibitemShut
  {NoStop}%
\bibitem [{\citenamefont {Jennewein}\ \emph {et~al.}(2000)\citenamefont
  {Jennewein}, \citenamefont {Achleitner}, \citenamefont {Weihs}, \citenamefont
  {Weinfurter},\ and\ \citenamefont {Zeilinger}}]{Jennewein:2000ly}%
  \BibitemOpen
  \bibfield  {author} {\bibinfo {author} {\bibfnamefont {T.}~\bibnamefont
  {Jennewein}}, \bibinfo {author} {\bibfnamefont {U.}~\bibnamefont
  {Achleitner}}, \bibinfo {author} {\bibfnamefont {G.}~\bibnamefont {Weihs}},
  \bibinfo {author} {\bibfnamefont {H.}~\bibnamefont {Weinfurter}}, \ and\
  \bibinfo {author} {\bibfnamefont {A.}~\bibnamefont {Zeilinger}},\ }\href
  {\doibase 10.1063/1.1150518} {\bibfield  {journal} {\bibinfo  {journal}
  {Review of Scientific Instruments}\ }\textbf {\bibinfo {volume} {71}},\
  \bibinfo {pages} {1675} (\bibinfo {year} {2000})}\BibitemShut {NoStop}%
\bibitem [{\citenamefont {Shen}\ \emph {et~al.}(2010)\citenamefont {Shen},
  \citenamefont {Tian},\ and\ \citenamefont {Zou}}]{Shen:2010vn}%
  \BibitemOpen
  \bibfield  {author} {\bibinfo {author} {\bibfnamefont {Y.}~\bibnamefont
  {Shen}}, \bibinfo {author} {\bibfnamefont {L.}~\bibnamefont {Tian}}, \ and\
  \bibinfo {author} {\bibfnamefont {H.}~\bibnamefont {Zou}},\ }\href {\doibase
  10.1103/PhysRevA.81.063814} {\bibfield  {journal} {\bibinfo  {journal}
  {Physical Review A}\ }\textbf {\bibinfo {volume} {81}} (\bibinfo {year}
  {2010}),\ 10.1103/PhysRevA.81.063814}\BibitemShut {NoStop}%
\bibitem [{\citenamefont {Stip{\v{c}}evi{\'{c}}}\ and\ \citenamefont
  {Rogina}(2007)}]{Stipcevic:2007fk}%
  \BibitemOpen
  \bibfield  {author} {\bibinfo {author} {\bibfnamefont {M.}~\bibnamefont
  {Stip{\v{c}}evi{\'{c}}}}\ and\ \bibinfo {author} {\bibfnamefont {B.~M.}\
  \bibnamefont {Rogina}},\ }\href {\doibase 10.1063/1.2720728} {\bibfield
  {journal} {\bibinfo  {journal} {Review of Scientific Instruments}\ }\textbf
  {\bibinfo {volume} {78}},\ \bibinfo {pages} {045104} (\bibinfo {year}
  {2007})}\BibitemShut {NoStop}%
\bibitem [{\citenamefont {{ID Quantique}}()}]{quantis-idQuantique}%
  \BibitemOpen
  \bibfield  {author} {\bibinfo {author} {\bibnamefont {{ID Quantique}}},\
  }\href@noop {} {\enquote {\bibinfo {title} {{Quantis QRNG}},}\ }\bibinfo
  {note}
  {\url{https://www.idquantique.com/random-number-generation/}}\BibitemShut
  {NoStop}%
\bibitem [{\citenamefont {Ma}\ \emph {et~al.}(2016)\citenamefont {Ma},
  \citenamefont {Yuan}, \citenamefont {Cao}, \citenamefont {Qi},\ and\
  \citenamefont {Zhang}}]{Ma16}%
  \BibitemOpen
  \bibfield  {author} {\bibinfo {author} {\bibfnamefont {X.}~\bibnamefont
  {Ma}}, \bibinfo {author} {\bibfnamefont {X.}~\bibnamefont {Yuan}}, \bibinfo
  {author} {\bibfnamefont {Z.}~\bibnamefont {Cao}}, \bibinfo {author}
  {\bibfnamefont {B.}~\bibnamefont {Qi}}, \ and\ \bibinfo {author}
  {\bibfnamefont {Z.}~\bibnamefont {Zhang}},\ }\href {\doibase
  10.1038/npjqi.2016.21} {\bibfield  {journal} {\bibinfo  {journal} {npj
  Quantum Information}\ }\textbf {\bibinfo {volume} {2}},\ \bibinfo {pages}
  {16021} (\bibinfo {year} {2016})}\BibitemShut {NoStop}%
\bibitem [{\citenamefont {Himbeeck}\ \emph {et~al.}(2017)\citenamefont
  {Himbeeck}, \citenamefont {Woodhead}, \citenamefont {Cerf}, \citenamefont
  {Garc\'{i}a-Pat\'{o}n},\ and\ \citenamefont {Pironio}}]{Himbeeck17}%
  \BibitemOpen
  \bibfield  {author} {\bibinfo {author} {\bibfnamefont {T.~V.}\ \bibnamefont
  {Himbeeck}}, \bibinfo {author} {\bibfnamefont {E.}~\bibnamefont {Woodhead}},
  \bibinfo {author} {\bibfnamefont {N.~J.}\ \bibnamefont {Cerf}}, \bibinfo
  {author} {\bibfnamefont {R.}~\bibnamefont {Garc\'{i}a-Pat\'{o}n}}, \ and\
  \bibinfo {author} {\bibfnamefont {S.}~\bibnamefont {Pironio}},\ }\href
  {\doibase 10.22331/q-2017-11-18-33} {\bibfield  {journal} {\bibinfo
  {journal} {Quantum}\ }\textbf {\bibinfo {volume} {1}},\ \bibinfo {pages} {33}
  (\bibinfo {year} {2017})}\BibitemShut {NoStop}%
\bibitem [{\citenamefont {Klyachko}\ \emph {et~al.}(2008)\citenamefont
  {Klyachko}, \citenamefont {Can}, \citenamefont {Binicio\v{g}ly},\ and\
  \citenamefont {Shumovsky}}]{Klyachko08}%
  \BibitemOpen
  \bibfield  {author} {\bibinfo {author} {\bibfnamefont {A.~A.}\ \bibnamefont
  {Klyachko}}, \bibinfo {author} {\bibfnamefont {M.~A.}\ \bibnamefont {Can}},
  \bibinfo {author} {\bibfnamefont {S.}~\bibnamefont {Binicio\v{g}ly}}, \ and\
  \bibinfo {author} {\bibfnamefont {A.~S.}\ \bibnamefont {Shumovsky}},\ }\href
  {\doibase 10.1103/PhysRevLett.101.020403} {\bibfield  {journal} {\bibinfo
  {journal} {Physical Review Letters}\ }\textbf {\bibinfo {volume} {101}},\
  \bibinfo {pages} {020403} (\bibinfo {year} {2008})}\BibitemShut {NoStop}%
\bibitem [{\citenamefont {Cabello}(2008)}]{Cabello08}%
  \BibitemOpen
  \bibfield  {author} {\bibinfo {author} {\bibfnamefont {A.}~\bibnamefont
  {Cabello}},\ }\href {\doibase 10.1103/PhysRevLett.101.210401} {\bibfield
  {journal} {\bibinfo  {journal} {Physical Review Letters}\ }\textbf {\bibinfo
  {volume} {101}},\ \bibinfo {pages} {210401} (\bibinfo {year}
  {2008})}\BibitemShut {NoStop}%
\bibitem [{\citenamefont {Um}\ \emph {et~al.}(2013)\citenamefont {Um},
  \citenamefont {Zhang}, \citenamefont {Zhang}, \citenamefont {Wang},
  \citenamefont {Yangchao}, \citenamefont {Deng}, \citenamefont {Duan},\ and\
  \citenamefont {Kim}}]{Um:2013tt}%
  \BibitemOpen
  \bibfield  {author} {\bibinfo {author} {\bibfnamefont {M.}~\bibnamefont
  {Um}}, \bibinfo {author} {\bibfnamefont {X.}~\bibnamefont {Zhang}}, \bibinfo
  {author} {\bibfnamefont {J.}~\bibnamefont {Zhang}}, \bibinfo {author}
  {\bibfnamefont {Y.}~\bibnamefont {Wang}}, \bibinfo {author} {\bibfnamefont
  {S.}~\bibnamefont {Yangchao}}, \bibinfo {author} {\bibfnamefont {D.-L.}\
  \bibnamefont {Deng}}, \bibinfo {author} {\bibfnamefont {L.-M.}\ \bibnamefont
  {Duan}}, \ and\ \bibinfo {author} {\bibfnamefont {K.}~\bibnamefont {Kim}},\
  }\href {\doibase 10.1038/srep01627} {\bibfield  {journal} {\bibinfo
  {journal} {Scientific Reports}\ }\textbf {\bibinfo {volume} {3}},\ \bibinfo
  {pages} {1627} (\bibinfo {year} {2013})}\BibitemShut {NoStop}%
\bibitem [{\citenamefont {Deng}\ \emph {et~al.}(2013)\citenamefont {Deng},
  \citenamefont {Zu}, \citenamefont {Chang}, \citenamefont {Hou}, \citenamefont
  {Yang}, \citenamefont {Wang},\ and\ \citenamefont {Duan}}]{Deng13}%
  \BibitemOpen
  \bibfield  {author} {\bibinfo {author} {\bibfnamefont {D.-L.}\ \bibnamefont
  {Deng}}, \bibinfo {author} {\bibfnamefont {C.}~\bibnamefont {Zu}}, \bibinfo
  {author} {\bibfnamefont {X.-Y.}\ \bibnamefont {Chang}}, \bibinfo {author}
  {\bibfnamefont {P.-Y.}\ \bibnamefont {Hou}}, \bibinfo {author} {\bibfnamefont
  {H.-X.}\ \bibnamefont {Yang}}, \bibinfo {author} {\bibfnamefont {Y.-X.}\
  \bibnamefont {Wang}}, \ and\ \bibinfo {author} {\bibfnamefont {L.-M.}\
  \bibnamefont {Duan}},\ }\href@noop {} {\  (\bibinfo {year} {2013})},\ \Eprint
  {http://arxiv.org/abs/1301.5364}{arXiv:1301.5364 [quant-ph]}\BibitemShut
  {NoStop}%
\bibitem [{\citenamefont {Miller}\ and\ \citenamefont {Shi}(2017)}]{Miller17}%
  \BibitemOpen
  \bibfield  {author} {\bibinfo {author} {\bibfnamefont {C.~A.}\ \bibnamefont
  {Miller}}\ and\ \bibinfo {author} {\bibfnamefont {Y.}~\bibnamefont {Shi}},\
  }\href {\doibase 10.1137/15M1044333} {\bibfield  {journal} {\bibinfo
  {journal} {SIAM Journal on Computing}\ }\textbf {\bibinfo {volume} {46}},\
  \bibinfo {pages} {1304} (\bibinfo {year} {2017})}\BibitemShut {NoStop}%
\bibitem [{\citenamefont {Marsaglia}(2005)}]{Marsaglia05}%
  \BibitemOpen
  \bibfield  {author} {\bibinfo {author} {\bibfnamefont {G.}~\bibnamefont
  {Marsaglia}},\ }\href@noop {} {\bibfield  {journal} {\bibinfo  {journal}
  {Interstat}\ }\textbf {\bibinfo {volume} {10}},\ \bibinfo {pages} {1}
  (\bibinfo {year} {2005})}\BibitemShut {NoStop}%
\bibitem [{\citenamefont {Wagon}(2004)}]{Wagon04}%
  \BibitemOpen
  \bibfield  {author} {\bibinfo {author} {\bibfnamefont {S.}~\bibnamefont
  {Wagon}},\ }in\ \href {\doibase 10.1007/978-1-4757-3240-5_58} {\emph
  {\bibinfo {booktitle} {Pi: A Source Book}}},\ \bibinfo {editor} {edited by\
  \bibinfo {editor} {\bibfnamefont {J.~L.}\ \bibnamefont {Berggren}}, \bibinfo
  {editor} {\bibfnamefont {J.~M.}\ \bibnamefont {Borwein}}, \ and\ \bibinfo
  {editor} {\bibfnamefont {P.~B.}\ \bibnamefont {Borwein}}}\ (\bibinfo
  {publisher} {Springer-Verlag},\ \bibinfo {address} {New York},\ \bibinfo
  {year} {2004})\ pp.\ \bibinfo {pages} {557--559}\BibitemShut {NoStop}%
\bibitem [{\citenamefont {Bailey}\ \emph {et~al.}(2012)\citenamefont {Bailey},
  \citenamefont {Borwein}, \citenamefont {Calude}, \citenamefont {Dinneen},
  \citenamefont {Dumitrescu},\ and\ \citenamefont {Yee}}]{MR3004253}%
  \BibitemOpen
  \bibfield  {author} {\bibinfo {author} {\bibfnamefont {D.~H.}\ \bibnamefont
  {Bailey}}, \bibinfo {author} {\bibfnamefont {J.~M.}\ \bibnamefont {Borwein}},
  \bibinfo {author} {\bibfnamefont {C.~S.}\ \bibnamefont {Calude}}, \bibinfo
  {author} {\bibfnamefont {M.~J.}\ \bibnamefont {Dinneen}}, \bibinfo {author}
  {\bibfnamefont {M.}~\bibnamefont {Dumitrescu}}, \ and\ \bibinfo {author}
  {\bibfnamefont {A.}~\bibnamefont {Yee}},\ }\href {\doibase
  10.1080/10586458.2012.665333} {\bibfield  {journal} {\bibinfo  {journal}
  {Experimental Mathematics}\ }\textbf {\bibinfo {volume} {21}},\ \bibinfo
  {pages} {375} (\bibinfo {year} {2012})}\BibitemShut {NoStop}%
\bibitem [{\citenamefont {Abbott}\ \emph
  {et~al.}(2014{\natexlab{b}})\citenamefont {Abbott}, \citenamefont
  {Bienvenu},\ and\ \citenamefont {Senno}}]{Abbott:2014rt}%
  \BibitemOpen
  \bibfield  {author} {\bibinfo {author} {\bibfnamefont {A.~A.}\ \bibnamefont
  {Abbott}}, \bibinfo {author} {\bibfnamefont {L.}~\bibnamefont {Bienvenu}}, \
  and\ \bibinfo {author} {\bibfnamefont {G.}~\bibnamefont {Senno}},\
  }\href@noop {} {\emph {\bibinfo {title} {Non-uniformity in the {Q}uantis
  Random Number Generator}}},\ \bibinfo {type} {{\textit{{CDMTCS} Research
  Report Series 472}}}\ (\bibinfo  {institution} {Centre for Discrete
  Mathematics and Theoretical Computer Science, University of Auckland},\
  \bibinfo {year} {2014})\BibitemShut {NoStop}%
\bibitem [{\citenamefont {Peres}(1992)}]{Peres:1992aa}%
  \BibitemOpen
  \bibfield  {author} {\bibinfo {author} {\bibfnamefont {Y.}~\bibnamefont
  {Peres}},\ }\href@noop {} {\bibfield  {journal} {\bibinfo  {journal} {The
  Annals of Statistics}\ }\textbf {\bibinfo {volume} {20}},\ \bibinfo {pages}
  {590} (\bibinfo {year} {1992})}\BibitemShut {NoStop}%
\bibitem [{\citenamefont {Abbott}\ and\ \citenamefont
  {Calude}(2012)}]{Abbott:2010ij}%
  \BibitemOpen
  \bibfield  {author} {\bibinfo {author} {\bibfnamefont {A.~A.}\ \bibnamefont
  {Abbott}}\ and\ \bibinfo {author} {\bibfnamefont {C.~S.}\ \bibnamefont
  {Calude}},\ }\href {\doibase 10.3233/COM-2012-001} {\bibfield  {journal}
  {\bibinfo  {journal} {Computability}\ }\textbf {\bibinfo {volume} {1}},\
  \bibinfo {pages} {59} (\bibinfo {year} {2012})}\BibitemShut {NoStop}%
\bibitem [{\citenamefont {Kovalsky}\ \emph {et~al.}(2018)\citenamefont
  {Kovalsky}, \citenamefont {Hnilo},\ and\ \citenamefont
  {Ag\"{u}ero}}]{Kovalsky18}%
  \BibitemOpen
  \bibfield  {author} {\bibinfo {author} {\bibfnamefont {M.~G.}\ \bibnamefont
  {Kovalsky}}, \bibinfo {author} {\bibfnamefont {A.~A.}\ \bibnamefont {Hnilo}},
  \ and\ \bibinfo {author} {\bibfnamefont {M.~B.}\ \bibnamefont {Ag\"{u}ero}},\
  }\href@noop {} {\  (\bibinfo {year} {2018})},\ \Eprint
  {http://arxiv.org/abs/1805.07161}{arXiv:1805.07161 [quant-ph]}\BibitemShut
  {NoStop}%
\bibitem [{\citenamefont {Matsumoto}\ and\ \citenamefont
  {Nishimura}(1998)}]{Matsumoto98}%
  \BibitemOpen
  \bibfield  {author} {\bibinfo {author} {\bibfnamefont {M.}~\bibnamefont
  {Matsumoto}}\ and\ \bibinfo {author} {\bibfnamefont {T.}~\bibnamefont
  {Nishimura}},\ }\href {\doibase 10.1145/272991.272995} {\bibfield  {journal}
  {\bibinfo  {journal} {ACM Transactions on Modeling and Computer Simulation}\
  }\textbf {\bibinfo {volume} {8}},\ \bibinfo {pages} {3} (\bibinfo {year}
  {1998})}\BibitemShut {NoStop}%
\bibitem [{\citenamefont {Salmon}\ \emph {et~al.}(2011)\citenamefont {Salmon},
  \citenamefont {Moraes}, \citenamefont {Dror},\ and\ \citenamefont
  {Shaw}}]{Random123}%
  \BibitemOpen
  \bibfield  {author} {\bibinfo {author} {\bibfnamefont {J.~K.}\ \bibnamefont
  {Salmon}}, \bibinfo {author} {\bibfnamefont {M.~A.}\ \bibnamefont {Moraes}},
  \bibinfo {author} {\bibfnamefont {R.~O.}\ \bibnamefont {Dror}}, \ and\
  \bibinfo {author} {\bibfnamefont {D.~E.}\ \bibnamefont {Shaw}},\ }in\ \href
  {\doibase 10.1145/2063384.2063405} {\emph {\bibinfo {booktitle} {Proceedings
  of the International Conference for High Performance Computing, Networking,
  Storage and Analysis (SC11)}}}\ (\bibinfo  {publisher} {ACM},\ \bibinfo
  {address} {New York},\ \bibinfo {year} {2011})\BibitemShut {NoStop}%
\bibitem [{\citenamefont {O'Neill}(2014)}]{PCG}%
  \BibitemOpen
  \bibfield  {author} {\bibinfo {author} {\bibfnamefont {M.~E.}\ \bibnamefont
  {O'Neill}},\ }\href {https://www.cs.hmc.edu/tr/hmc-cs-2014-0905.pdf} {\emph
  {\bibinfo {title} {PCG: A Family of Simple Fast Space-Efficient Statistically
  Good Algorithms for Random Number Generation}}},\ \bibinfo {type} {Tech.
  Rep.}\ \bibinfo {number} {HMC-CS-2014-0905}\ (\bibinfo  {institution} {Harvey
  Mudd College},\ \bibinfo {address} {Claremont, CA},\ \bibinfo {year}
  {2014})\BibitemShut {NoStop}%
\bibitem [{\citenamefont {Marsaglia}(2003)}]{xoroshift}%
  \BibitemOpen
  \bibfield  {author} {\bibinfo {author} {\bibfnamefont {G.}~\bibnamefont
  {Marsaglia}},\ }\href {\doibase 10.18637/jss.v008.i14} {\bibfield  {journal}
  {\bibinfo  {journal} {Journal of Statistical Software}\ }\textbf {\bibinfo
  {volume} {8}} (\bibinfo {year} {2003}),\ 10.18637/jss.v008.i14}\BibitemShut
  {NoStop}%
\bibitem [{\citenamefont {Solis}\ \emph {et~al.}(2015)\citenamefont {Solis},
  \citenamefont {Mart\'{i}nez}, \citenamefont {Alarc\'{o}n}, \citenamefont
  {Ram\'{i}rez}, \citenamefont {U'Ren},\ and\ \citenamefont
  {Hirsch}}]{Solis15}%
  \BibitemOpen
  \bibfield  {author} {\bibinfo {author} {\bibfnamefont {A.}~\bibnamefont
  {Solis}}, \bibinfo {author} {\bibfnamefont {A.~M.~A.}\ \bibnamefont
  {Mart\'{i}nez}}, \bibinfo {author} {\bibfnamefont {R.~R.}\ \bibnamefont
  {Alarc\'{o}n}}, \bibinfo {author} {\bibfnamefont {H.~C.}\ \bibnamefont
  {Ram\'{i}rez}}, \bibinfo {author} {\bibfnamefont {A.~B.}\ \bibnamefont
  {U'Ren}}, \ and\ \bibinfo {author} {\bibfnamefont {J.~G.}\ \bibnamefont
  {Hirsch}},\ }\href {\doibase 10.1088/0031-8949/90/7/074034} {\bibfield
  {journal} {\bibinfo  {journal} {Physica Scripta}\ }\textbf {\bibinfo {volume}
  {90}},\ \bibinfo {pages} {074034} (\bibinfo {year} {2015})}\BibitemShut
  {NoStop}%
\bibitem [{\citenamefont {Mart\'{i}nez}\ \emph {et~al.}(2018)\citenamefont
  {Mart\'{i}nez}, \citenamefont {Sol\'{i}s}, \citenamefont {Rojas},
  \citenamefont {U'Ren}, \citenamefont {Hirsch},\ and\ \citenamefont
  {Castillo}}]{Martinez18}%
  \BibitemOpen
  \bibfield  {author} {\bibinfo {author} {\bibfnamefont {A.~C.}\ \bibnamefont
  {Mart\'{i}nez}}, \bibinfo {author} {\bibfnamefont {A.}~\bibnamefont
  {Sol\'{i}s}}, \bibinfo {author} {\bibfnamefont {R.~D.~H.}\ \bibnamefont
  {Rojas}}, \bibinfo {author} {\bibfnamefont {A.~B.}\ \bibnamefont {U'Ren}},
  \bibinfo {author} {\bibfnamefont {J.~G.}\ \bibnamefont {Hirsch}}, \ and\
  \bibinfo {author} {\bibfnamefont {I.~P.}\ \bibnamefont {Castillo}},\
  }\href@noop {} {\  (\bibinfo {year} {2018})},\ \Eprint
  {http://arxiv.org/abs/1810.08718}{arXiv:1810.08718 [quant-ph]}\BibitemShut
  {NoStop}%
\bibitem [{\citenamefont {Conover}(1999)}]{Conover}%
  \BibitemOpen
  \bibfield  {author} {\bibinfo {author} {\bibfnamefont {W.~J.}\ \bibnamefont
  {Conover}},\ }\href@noop {} {\emph {\bibinfo {title} {Practical Nonparametric
  Statistics}}}\ (\bibinfo  {publisher} {John Wiley \& Sons},\ \bibinfo
  {address} {New York},\ \bibinfo {year} {1999})\ p.\ \bibinfo {pages}
  {584}\BibitemShut {NoStop}%
\bibitem [{\citenamefont {Shapiro}\ and\ \citenamefont
  {Wilk}(2005)}]{Shapiro-Wilk}%
  \BibitemOpen
  \bibfield  {author} {\bibinfo {author} {\bibfnamefont {S.~S.}\ \bibnamefont
  {Shapiro}}\ and\ \bibinfo {author} {\bibfnamefont {M.~B.}\ \bibnamefont
  {Wilk}},\ }\href {\doibase 10.1093/biomet/52.3-4.591} {\bibfield  {journal}
  {\bibinfo  {journal} {Biometrika}\ }\textbf {\bibinfo {volume} {52}},\
  \bibinfo {pages} {591} (\bibinfo {year} {2005})}\BibitemShut {NoStop}%
\bibitem [{\citenamefont {Welch}(1947)}]{Welch}%
  \BibitemOpen
  \bibfield  {author} {\bibinfo {author} {\bibfnamefont {B.~L.}\ \bibnamefont
  {Welch}},\ }\href {\doibase 10.1093/biomet/34.1-2.28} {\bibfield  {journal}
  {\bibinfo  {journal} {Biometrika}\ }\textbf {\bibinfo {volume} {34}},\
  \bibinfo {pages} {28} (\bibinfo {year} {1947})}\BibitemShut {NoStop}%
\bibitem [{\citenamefont {Pinch}(2007)}]{Pinch07}%
  \BibitemOpen
  \bibfield  {author} {\bibinfo {author} {\bibfnamefont {R.~G.~E.}\
  \bibnamefont {Pinch}},\ }in\ \href@noop {} {\emph {\bibinfo {booktitle}
  {Proceedings of Conference on Algorithmic Number Theory 2007}}},\
  Vol.~\bibinfo {volume} {46},\ \bibinfo {editor} {edited by\ \bibinfo {editor}
  {\bibfnamefont {A.-M.}\ \bibnamefont {Ernvall-Hyt{\"{o}}nen}}, \bibinfo
  {editor} {\bibfnamefont {M.}~\bibnamefont {Jutila}}, \bibinfo {editor}
  {\bibnamefont {Juhani}}, \bibinfo {editor} {\bibnamefont {Karhum{\"{a}}ki}},
  \ and\ \bibinfo {editor} {\bibfnamefont {A.}~\bibnamefont {Lepist{\"{o}}}}}\
  (\bibinfo {year} {2007})\ pp.\ \bibinfo {pages} {129--131}\BibitemShut
  {NoStop}%
\bibitem [{\citenamefont {Abbott}\ \emph {et~al.}(2018)\citenamefont {Abbott},
  \citenamefont {Calude}, \citenamefont {Dinneen},\ and\ \citenamefont
  {Huang}}]{CDMTCS515V2}%
  \BibitemOpen
  \bibfield  {author} {\bibinfo {author} {\bibfnamefont {A.~A.}\ \bibnamefont
  {Abbott}}, \bibinfo {author} {\bibfnamefont {C.~S.}\ \bibnamefont {Calude}},
  \bibinfo {author} {\bibfnamefont {M.~J.}\ \bibnamefont {Dinneen}}, \ and\
  \bibinfo {author} {\bibfnamefont {N.}~\bibnamefont {Huang}},\ }\href
  {https://www.cs.auckland.ac.nz/research/groups/CDMTCS/researchreports/index.php?download&data_file=13}
  {\emph {\bibinfo {title} {Experimental Probing of the Incomputability of
  Quantum Randomness}}},\ \bibinfo {type} {{\textit{{CDMTCS} Research Report
  Series 515v2}}}\ (\bibinfo  {institution} {Centre for Discrete Mathematics
  and Theoretical Computer Science, University of Auckland},\ \bibinfo {year}
  {2018})\ \bibinfo {note}
  {\url{https://www.cs.auckland.ac.nz/research/groups/CDMTCS/export/80_random_seqs/}}\BibitemShut
  {NoStop}%
\bibitem [{\citenamefont {Solovay}\ and\ \citenamefont
  {Strassen}(1977{\natexlab{b}})}]{Solovay77b}%
  \BibitemOpen
  \bibfield  {author} {\bibinfo {author} {\bibfnamefont {R.}~\bibnamefont
  {Solovay}}\ and\ \bibinfo {author} {\bibfnamefont {V.}~\bibnamefont
  {Strassen}},\ }\href {\doibase 10.1137/0207009} {\bibfield  {journal}
  {\bibinfo  {journal} {SIAM Journal on Computing}\ }\textbf {\bibinfo {volume}
  {7}},\ \bibinfo {pages} {118} (\bibinfo {year}
  {1977}{\natexlab{b}})}\BibitemShut {NoStop}%
\end{thebibliography}%

\appendix

\clearpage

\section{Chaitin-Schwartz-Solovay-Strassen test analysis tables}
\label{app:CSSS1tables}

\renewcommand{\theequation}{A\arabic{equation}}
\setcounter{equation}{0}

\renewcommand{\thetable}{A\arabic{table}}
\setcounter{table}{0}

\renewcommand{\thefigure}{A\arabic{figure}}
\setcounter{figure}{0}

\begin{table}[ht]
\begin{center}
\caption{Kolmogorov-Smirnov tests for the first Chaitin-Schwartz-Solovay-Strassen test with the metric that records the minimum number
	of witnesses needed to verify the compositeness of all Carmichael numbers of at most 16 digits.}\label{tab:CSSStest1CW16KS}
\medskip
\begin{tabular}{cccccc}
	\hline
	\hline
	$p$-values & $\pi$ & Python & Random123 & QRNG & xoroshiro128+\\
	\hline\\[-2ex]
	PCG        &  0.8186 & 0.8186 & 1  & 1 & 1 \\ 
	$\pi$      &       & 0.9976 & 0.9976   & 0.5596 & 1 \\
	Python     &	   &	   & 0.9976   & 0.5596 & 0.9976 \\
	Random123  &	   & 	   & 		 & 0.9976 & 1 \\
	QRNG       &	   & 	   & 		 &		 & 0.9780 \\
	\hline
	\hline
\end{tabular}
\end{center}
\end{table}

\begin{table}[ht]
\begin{center}
\caption{Shapiro-Wilk tests of normality for the first Chaitin-Schwartz-Solovay-Strassen test with the metric that records the minimum number
	of witnesses needed to verify the compositeness of all Carmichael numbers of at most 16 digits.}\label{tab:CSSStest1CW16SW}
\medskip
\centering
\begin{tabular}{ccccccc}
	\hline
	\hline
	& PCG  & $\pi$ & Python & Random123 & QRNG & xoroshiro128+\\
	\hline\\[-2ex]
				$p$-value  & $\mathbf{< 10^{-4}}$  & $\mathbf{< 10^{-4}}$ & $\mathbf{< 10^{-4}}$ & $\mathbf{< 10^{-4}}$ & $\mathbf{< 10^{-4}}$ & $\mathbf{< 10^{-4}}$ \\ 
	\hline
	\hline
\end{tabular}
\end{center}
\end{table}

\begin{table}[ht]
	\begin{center}
		\caption{Kolmogorov-Smirnov tests for the second Chaitin-Schwartz-Solovay-Strassen test with the ``bit counting'' metric on the non-complemented (i.e., original) bits.}\label{tab:CSSStest1CbOrKS}
		\medskip
		\begin{tabular}{cccccc}
			\hline
			\hline
			$p$-values & $\pi$ & Python & Random123 & QRNG & xoroshiro128+\\
			\hline\\[-2ex]
			PCG        & 0.6953 & 0.4383 & 0.922   & 0.0132 & 0.6953 \\ 
			$\pi$      &       & 0.4383 & 0.8219   & \textbf{0.0045} & 0.9794 \\
			Python     &	   &	   & 0.0814   & 0.0537 & 0.5625 \\
			Random123  &	   & 	   & 		 & \textbf{0.0014} & 0.5625 \\
			QRNG       &	   & 	   & 		 &		 & \textbf{0.0026} \\
			\hline
			\hline
		\end{tabular}
	\end{center}
\end{table}

\begin{table}[ht]
	\begin{center}
		\caption{Shapiro-Wilk tests of normality for the second Chaitin-Schwartz-Solovay-Strassen test with the ``bit counting'' metric on the non-complemented (i.e., original) bits.}\label{tab:CSSStest1CbOrSW}
		\medskip
		\begin{tabular}{ccccccc}
			\hline
			\hline
			& PCG  & $\pi$ & Python & Random123 & QRNG & xoroshiro128+\\
			\hline\\[-2ex]
			$p$-value  & 0.4892  & 0.2003 & \textbf{0.04867} & 0.5951 & 0.1669 & 0.0808 \\ 
			\hline
			\hline
		\end{tabular}
	\end{center}
\end{table}

\begin{table}[ht]
\begin{center}
\caption{Kolmogorov-Smirnov tests for the second Chaitin-Schwartz-Solovay-Strassen test with the ``bit counting'' metric on the complemented bits.}\label{tab:CSSStest1CbCoKS}
\medskip
\begin{tabular}{cccccc}
	\hline
	\hline
	$p$-values & $\pi$ & Python & Random123 & QRNG & xoroshiro128+\\
	\hline\\[-2ex]
	PCG        & 0.4383 & 0.3307 & 0.2424   & 0.05372 & 0.5625 \\ 
	$\pi$      &       & 0.4383 & 0.1202   & \textbf{0.0045} & 0.5625 \\
	Python     &	   &	   & 0.5625   & \textbf{0.0026} & 0.8219 \\
	Random123  &	   & 	   & 		 & \textbf{0.0014} & 0.2424 \\
	QRNG       &	   & 	   & 		 &		 & 0.0132 \\
	\hline
	\hline
\end{tabular}
\end{center}
\end{table}

\begin{table}[ht]
\begin{center}
\caption{Shapiro-Wilk tests of normality for the second Chaitin-Schwartz-Solovay-Strassen test with the ``bit counting'' metric on the complemented bits.}\label{tab:CSSStest1CbCoSW}
\medskip
\begin{tabular}{ccccccc}
	\hline
	\hline
	& PCG  & $\pi$ & Python & Random123 & QRNG & xoroshiro128+\\
	\hline\\[-2ex]
	$p$-value  & 0.199 & 0.2433 & 0.0754 & 0.4401 & 0.0518 & 0.9673 \\
	\hline
	\hline
\end{tabular}
\end{center}
\end{table}

\begin{table}[ht]
\begin{center}
\caption{Welch $t$-tests for the second Chaitin-Schwartz-Solovay-Strassen test with the ``bit counting'' metric on the complemented bits.}\label{tab:CSSStest1CbCoWt}
\medskip
\begin{tabular}{cccccc}
	\hline
	\hline
	$p$-values & $\pi$ & Python & Random123 & QRNG & xoroshiro128+\\
	\hline\\[-2ex]
	PCG        & 0.6422 & 0.3796 & 0.1265   &  \textbf{0.0034} & 0.9454 \\
	$\pi$      &       & 0.6343 & 0.2287   & \textbf{0.0004} & 0.6795 \\
	Python     &	   &	   & 0.4683   & \textbf{0.0001} & 0.3964 \\
	Random123  &	   & 	   & 		 & $\mathbf{<10^{-4}}$ & 0.1271 \\
	QRNG       &	   & 	   & 		 &		 & \textbf{0.0020} \\
	\hline
	\hline
\end{tabular}
\end{center}
\end{table}

\begin{table}[ht]
	\begin{center}
		\caption{Kolmogorov-Smirnov tests for the third Chaitin-Schwartz-Solovay-Strassen test with the ``bit-counting'' metric for the non-complemented (i.e., original) bits for all Carmichael numbers of at most 16 digits.}\label{tab:CSSS2M1_KS}
		\medskip
		\begin{tabular}{cccccc}
			\hline
			\hline
			$p$-values & $\pi$ & Python & Random123 & QRNG & xoroshiro128+\\
			\hline\\[-2ex]
			PCG        & 0.2694 & 0.4821 & 0.2988   & 0.4013 & 0.1054 \\ 
			$\pi$      &       & 0.6953 & 0.4383   & 0.3307 & 0.4383 \\
			Python     &	   &	   & 0.8186   & 0.5625 & 0.5625 \\
			Random123  &	   & 	   & 		 & 0.9794 & 0.8219 \\
			QRNG       &	   & 	   & 		 &		 & 0.8219 \\
			\hline
			\hline
		\end{tabular}
	\end{center}
\end{table}

\begin{table}[ht]
	\begin{center}
		\caption{Shapiro-Wilk tests of normality for the third Chaitin-Schwartz-Solovay-Strassen test with the ``bit-counting'' metric for the non-complemented (i.e., original) bits for all Carmichael numbers of at most 16 digits.}\label{tab:CSSS2M1_SW}
		\medskip
		\begin{tabular}{ccccccc}
			\hline
			\hline
			 		   & PCG  & $\pi$ & Python & Random123 & QRNG & xoroshiro128+\\
			\hline\\[-2ex]
			$p$-value  & 0.2076  & 0.4921 & 0.3337 & 0.1956 & 0.7608 & 0.1347 \\ 
			\hline
			\hline
		\end{tabular}
	\end{center}
\end{table}

\begin{table}[ht]
	\begin{center}
		\caption{Welch $t$-tests for the third Chaitin-Schwartz-Solovay-Strassen test with the ``bit-counting'' metric for the non-complemented (i.e., original) bits for all Carmichael numbers of at most 16 digits.}\label{tab:CSSS2M1_Wt}
		\medskip
		\begin{tabular}{cccccc}
			\hline
			\hline
			$p$-values & $\pi$ & Python & Random123 & QRNG & xoroshiro128+\\
			\hline
			PCG        & 0.2838 & 0.81 & 0.5227   & 0.4335 & 0.2437 \\ 
			$\pi$      &       & 0.4186 & 0.6833   & 0.8401 & 0.911 \\
			Python     &	   &	   & 0.6956   & 0.584 & 0.3653 \\
			Random123  &	   & 	   & 		 & 0.8585 & 0.6096 \\
			QRNG       &	   & 	   & 		 &		 & 0.7629 \\
			\hline
			\hline
		\end{tabular}
	\end{center}
\end{table}

\begin{table}[ht]
	\begin{center}
		\caption{Kolmogorov-Smirnov tests for the third Chaitin-Schwartz-Solovay-Strassen test with the ``bit-counting'' metric for the complemented bits for all Carmichael numbers of at most 16 digits.}\label{tab:CSSS2M1C_KS}
		\medskip
		\begin{tabular}{cccccc}
			\hline
			\hline
			$p$-values & $\pi$ & Python & Random123 & QRNG & xoroshiro128+\\
			\hline
			PCG        & 0.5596 & 0.9794 & 0.173   & 0.9794 & 0.3307 \\ 
			$\pi$      &       & 0.922 & 0.8219   & 0.8219 & 0.6953 \\
			Python     &	   &	   & 0.5625   & 0.9194 & 0.6953 \\
			Random123  &	   & 	   & 		 & 0.4383 & 0.1201 \\
			QRNG       &	   & 	   & 		 &		 & 0.8219 \\
			\hline
			\hline
		\end{tabular}
	\end{center}
\end{table}

\begin{table}[ht]
	\begin{center}
		\caption{Shapiro-Wilk tests of normality for the third Chaitin-Schwartz-Solovay-Strassen test with the ``bit-counting'' metric for the complemented bits for all Carmichael numbers of at most 16 digits.}\label{tab:CSSS2M1C_SW}
		\medskip
		\begin{tabular}{ccccccc}
			\hline
			\hline
			& PCG  & $\pi$ & Python & Random123 & QRNG & xoroshiro128+\\
			\hline
			$p$-value  & 0.4616  & 0.6708 & 0.6067 & 0.94 & 0.9355 & \textbf{0.0239} \\ 
			\hline
			\hline
		\end{tabular}
	\end{center}
\end{table}

\begin{table}[ht]
	\begin{center}
		\caption{Kolmogorov-Smirnov tests for the fourth Chaitin-Schwartz-Solovay-Strassen test with the ``violation-count'' metric for non-complemented (i.e., original) bits for all odd composite numbers that are less than $50$.}\label{tab:CSSS2M2_KS}
		\medskip
		\begin{tabular}{cccccc}
			\hline
			\hline
			$p$-values & $\pi$ & Python & Random123 & QRNG & xoroshiro128+\\
			\hline
            PCG        & 0.318 & 0.2414 & 0.692   & \textbf{0.0027} & 0.9976 \\ 
			$\pi$      &       & 0.692 & 0.8186   & 0.05397 & 0.9976 \\
			Python     &	   &	   & 0.9194   & \textbf{0.0004} & 0.8186 \\
			Random123  &	   & 	   & 		 & \textbf{0.0047} & 0.8186 \\
			QRNG       &	   & 	   & 		 &		 & 0.0348 \\
			\hline
			\hline
		\end{tabular}
	\end{center}
\end{table}

\begin{table}[ht]
	\begin{center}
		\caption{Shapiro-Wilk tests of normality for the fourth Chaitin-Schwartz-Solovay-Strassen test with the ``violation-count'' metric for non-complemented (i.e., original) bits for all odd composite numbers that are less than $50$.}\label{tab:CSSS2M2_SW}
		\medskip
		\begin{tabular}{ccccccc}
			\hline
			\hline
			 		   & PCG  & $\pi$ & Python & Random123 & QRNG & xoroshiro128+\\
			\hline
			$p$-value  & $\mathbf{<10^{-4}}$  & \textbf{0.0040} & \textbf{0.0002} & \textbf{0.0056} & \textbf{0.0115} & \textbf{0.0148} \\ 
			\hline
			\hline
		\end{tabular}
	\end{center}
\end{table}

\clearpage

\begin{table}[ht]
\begin{center}
\caption{Kolmogorov-Smirnov tests for the fourth Chaitin-Schwartz-Solovay-Strassen test with the ``violation-count'' metric for the complemented bits for all odd composite numbers that are less than $50$.}\label{tab:CSSS2M2C_KS}
\medskip
\begin{tabular}{cccccc}
	\hline
	\hline
	$p$-values & $\pi$ & Python & Random123 & QRNG & xoroshiro128+\\
	\hline
	PCG        & 0.692 & 0.9194 & 0.9194 & 0.1725 & 0.5596 \\ 
	$\pi$      &       & 0.5596 & 0.9976   & 0.692 & 0.2414 \\
	Python     &	   &	   & 0.692   & 0.1725 & 0.8186 \\
	Random123  &	   & 	   & 		 & 0.5596 & 0.5596 \\
	QRNG       &	   & 	   & 		 &		 & 0.0135 \\
	\hline
	\hline
\end{tabular}
\end{center}
\end{table}

\begin{table}[ht]
\begin{center}
		\caption{Shapiro-Wilk tests of normality for the fourth Chaitin-Schwartz-Solovay-Strassen test with the ``violation-count'' metric for the complemented bits for all odd composite numbers that are less than $50$.}\label{tab:CSSS2M2C_SW}
		\medskip
        \centering
			\begin{tabular}{ccccccc}
				\hline
				\hline
				& PCG  & $\pi$ & Python & Random123 & QRNG & xoroshiro128+\\
				\hline
				$p$-value  & 0.06601 &  \textbf{0.02957} & $\mathbf{<10^{-4}}$ & \textbf{0.0080} & $\mathbf{<10^{-4}}$ & \textbf{0.0017} \\ 
				\hline
				\hline
			\end{tabular}
		\end{center}
	
\end{table}

\end{document}